\documentclass[twocolumn]{aastex631}

\usepackage{siunitx}
\usepackage{amsmath}
\usepackage{natbib}
\usepackage{hyperref}

\begin{document}

\newcommand{\Reply}[1]{{\color{cyan} \bf #1}}

\title{H$\alpha$ Time Delays of AGNs from the Zwicky Transcient Facility Broadband Photometry}

\email{maqinchun@pku.edu.cn; wuxb@pku.edu.cn}
\author{Qinchun Ma}
\affiliation{Department of Astronomy, School of Physics, Peking University, Beijing 100871, China}
\affiliation{Kavli Institute for Astronomy and Astrophysics, Peking University, Beijing 100871, China}

\author{Yuhan Wen}
\affiliation{Department of Astronomy, School of Physics, Peking University, Beijing 100871, China}
\affiliation{Kavli Institute for Astronomy and Astrophysics, Peking University, Beijing 100871, China}

\author{Xue-Bing Wu}
\affiliation{Department of Astronomy, School of Physics, Peking University, Beijing 100871, China}
\affiliation{Kavli Institute for Astronomy and Astrophysics, Peking University, Beijing 100871, China}

\author{Huapeng Gu}
\affiliation{Department of Astronomy, School of Physics, Peking University, Beijing 100871, China}
\affiliation{Kavli Institute for Astronomy and Astrophysics, Peking University, Beijing 100871, China}

\author{Yuming Fu}
\affiliation{Department of Astronomy, School of Physics, Peking University, Beijing 100871, China}
\affiliation{Kavli Institute for Astronomy and Astrophysics, Peking University, Beijing 100871, China}

\begin{abstract}

In our previous work on broadband photometric reverberation mapping (PRM), we proposed the ICCF-Cut process to obtain the time lags of H$\alpha$ emission line from two broadband lightcurves via subtracting the continuum emission from the line band. Extending the work, we enlarge our sample to the Zwicky Transient Facility (ZTF) database. We adopt two criteria to select 123 type 1 AGNs with sufficient variability and smooth lightcurves from 3537 AGNs at $z<0.09$ with more than 100 epoch observations in the $g$ and $r$ bands from the ZTF database. We calculate the H$\alpha$ time lags for 23 of them which have previous spectroscopic reverberation mapping (SRM) results using ICCF-Cut, JAVELIN and $\chi ^2$ methods. Our obtained H$\alpha$ time lags are slightly larger than the H$\beta$ time lags, which is consistent with the previous SRM results and the theoretical model of the AGN broad line region. The comparisons between SRM and PRM lag distributions and between the subtracted emission line lightcurves indicate that after selecting AGNs with the two criteria, combining the ICCF-Cut, JAVELIN and $\chi^2$ methods provides an efficient way to get the reliable H$\alpha$ lags from the broadband PRM. Such techniques can be used to estimate the black hole masses of a large sample of AGNs in the large multi-epoch photometric sky surveys such as the Legacy Survey of Space and Time (LSST) and the survey from the Wide Field Survey Telescope (WFST) in the near future.

\end{abstract}


\section{Introduction}

Reverberation mapping \citep[RM;][]{1982ApJ...255..419B,1993PASP..105..247P} 
is an efficient and widely used method to determine the masses of the supermassive black holes (SMBHs) in active galactic nuclei (AGNs). The principle of RM is that the photons from the accretion disk travel across the broad line region (BLR) and generate broad emission lines by photoionization process, so that the time lag $\tau$ between the optical continuum and broad emission line lightcurves reflects the size of the BLR, i.e. $R_{BLR}=c\tau$. By monitoring the continuum and the broad emission lines, we can obtain their lightcurves and calculate the time lag. By combining the velocity dispersion of the broad emission line in the spectra, we can calculate the virial mass of the SMBH at the center of the AGN. 

Spectroscopic RM (SRM) uses the medium- to large-sized telescopes to monitor the variability of the continuum and line emissions. It is usually very expensive and time-consuming. Although multi-fiber spectroscopic telescopes can obtain thousands of spectra in a single exposure, it is hard to achieve the high accuracy in the flux calibration required by the RM. Accurate fluxes are essential for extracting the lightcurves of the continuum and emission line, and calculating the time lag between them. Only several campaigns such as the Sloan Digital Sky Survey RM project \citep[SDSS RM;][]{2017ApJ...851...21G} and the Australian Dark Energy Survey (OzDES) RM program \citep{2021MNRAS.507.3771Y} use the fiber spectra, while most SRM campaigns use the slit spectra to monitor the AGN continuum and broad emission lines \citep[e.g.,][]{2009ApJ...705..199B, 2018ApJ...869..142D, 2022ApJS..262...14B}. In the past decades, only about 200 AGNs have the BLR sizes measured with SRM. To obtain the BLR sizes of a large sample of AGNs, a more efficient method is needed.

Photometric reverberation mapping (PRM) employs multi-band photometric observations to trace the AGN continuum and broad emission lines. The narrow and intermediate band PRMs use the specially designed filters with narrow bandwidth (FWHM around \SI{30}{\angstrom}) or intermediate bandwidth (FWHM around \SI{200}{\angstrom}) to collect photons mainly from the broad emission lines, while blocking most of the continuum emissions. For the sufficiently strong emission lines, such as H$\alpha$, lightcurves obtained from an appropriate narrow band can be directly considered as the emission line lightcurves, because the line component is dominant in this case \citep[e.g.,][]{2011A&A...535A..73H, 2018A&A...620A.137R}. For other cases, including the narrow band PRM for slightly weaker emission lines (such as H$\beta$) and the intermediate band PRM, previous works \citep[e.g.,][]{2012A&A...545A..84P,2016ApJ...818..137J} used a broad band which contains the narrow/intermediate band to determine the continuum component in the narrow/intermediate band. These methods have been proven effective by many campaigns. Nevertheless, the special narrow/intermediate band filters are only equipped on certain telescopes, and they also limit the redshift range and the sample size of AGNs. 

Another PRM technique, the broadband PRM, can be applied to a large sample of AGNs across a wide redshift range with the multi-epoch data from large photometric sky surveys. While compared to the narrow/intermediate band PRM, broadband PRM is more difficult for measuring the lags of broad emission lines, since the broadband photometric data contains mostly the continuum flux from the accretion disk rather than the emission lines from BLR. It is critical to remove the contribution of the continuum and extract the broad emission line from the broadband (this band is named as line band hereafter). Unlike the narrow/intermediate band cases, the continuum contribution is hard to determine in the line band of broadband PRM, because we need to use an adjacent broadband (here after continuum band) rather than the line band to trace the continuum. The different wavelength ranges and the small ratio of the emission line in the line band require that the continuum component should be carefully considered.

For broadband PRM, \citet{2012ApJ...747...62C} proposed the CCF-ACF (Cross-Correlation Function minus Auto-Correlation Function) method to subtract the continuum in the correlation domain, which assumes that the continuum component in the line band is identical to that in the continuum band. This method was adopted in some later works \citep[e.g.,][]{2012ApJ...756...73E,2013ApJ...773...24R}. \citet{2013A&A...552A...1P} presented the results of both narrow band PRM and broad band PRM (using the CCF-ACF method), which showed consistency of both techniques. However, the simple assumption of the continuum components in two bands being the same has large errors due to the AGN spectral slope and the varying flux ratio of the two continuum components in different AGNs. To improve this, further works \citep[e.g.,][]{2013ApJ...769..124C} considered the flux ratio of the emission line in the line band as an additional free parameter. This parameter adds to the degrees of freedom and is independent of the spectrum, so it may not match the real spectrum of the AGN.

In our previous work \citep[][hereafter Paper I]{2023ApJ...949...22M}, we proposed the ICCF-Cut process to remove the continuum emission and isolate the emission line from the broadband lightcurves. This procedure essentially follows the continuum removal proposed by \citet{2011A&A...535A..73H} and \citet{2012A&A...545A..84P} with some modifications. We also note that other techniques where the continuum is removed in the correlation domain have been proposed by \citet{2012ApJ...747...62C}. Compared with the CCF-ACF method, ICCF-Cut directly uses a single-epoch spectrum to determine the ratio and extract the emission line lightcurves. In Paper I, we selected 4 Seyfert 1 galaxies with simultaneous broadband and H$\beta$ lightcurves. We calculated their H$\alpha$ time lags using 2 broadband lightcurves and further compared our extracted H$\alpha$ lightcurves with the simultaneous H$\beta$ lightcurves to examine our results. This method, along with the JAVELIN and $\chi^2$ methods, proves to be an efficient way to measure reliable H$\alpha$ lags from broadband photometry. In this paper, we extend the range of AGN target selection for broadband PRM and apply these methods to the multi-epoch data from Zwicky Transient Facility (ZTF) \citep{2019PASP..131a8003M}.

ZTF is a time-domain survey that had its first light at Palomar Observatory in 2017. By scanning more than 3750 square degrees of the sky per hour to a depth of 20.5 mag, ZTF can produce a photometric variability catalog with nearly 300 observational epochs each year, which is ideal for the studies of the variabilities of AGN. These data can be used to improve the quality of the continuum lightcurves in SRM \citep{2021ApJS..253...20H} and to calculate the time lags between two broadbands for continuum RM \citep{2022MNRAS.511.3005J}. In spite of these advantages, we find that only the sources with large variabilities and high continuities in the lightcurves can serve as good candidates for the broadband PRM.
Therefore, in this work, we use the data of ZTF DR16 between March, 2018 and January, 2023, and adopt two the criteria to select AGNs that are good candidates for the broadband PRM. Same as Paper I, we use three methods, namely ICCF-Cut, JAVELIN and $\chi^2$ methods, to calculate the time lags to ensure the reliabilities of the time lags. To check our results further, we also compare the ZTF broadband PRM lags with the time lags obtained by SRM.

This paper is arranged as follows. We describe the target selections in Section 2. The time lags obtained using three methods are presented in Section 3. The discussion and the comparisons between PRM and SRM are presented in Section 4. A summary is given in Section 5.

\section{Target selections}

\begin{figure}[!h]
  \includegraphics[width=1.0\linewidth]{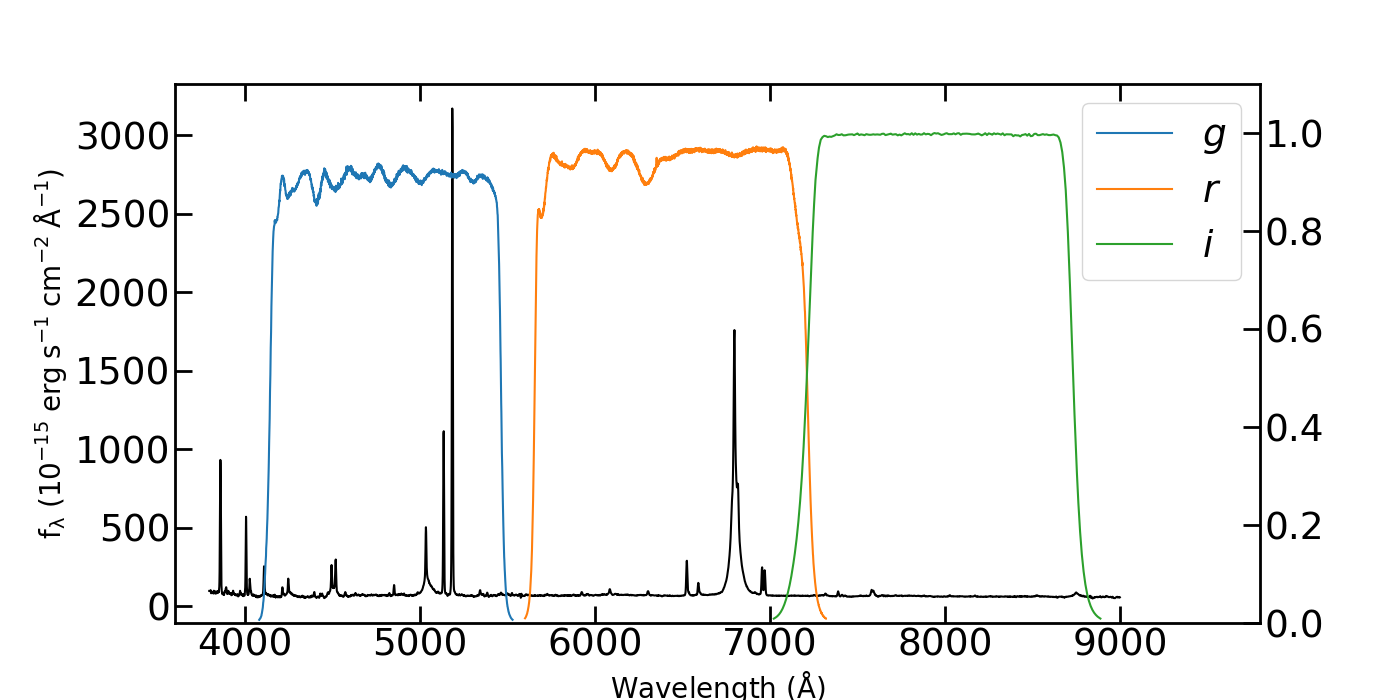}
  \caption{The spectrum of Mrk 110 and the transmission functions of the ZTF broad bands.}
\end{figure}

\begin{figure*}[!htb]
  \includegraphics[width=1.0\linewidth]{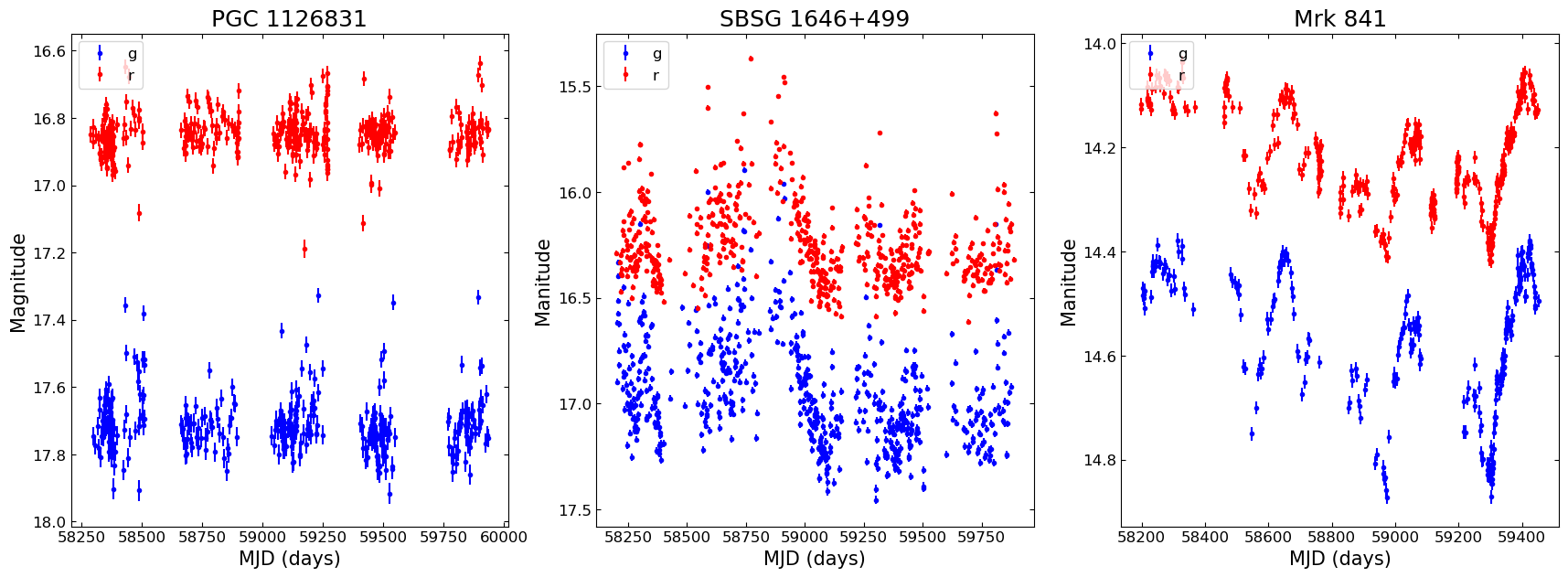}
  \caption{The example of one-day binned $g$ and $r$ band lightcurves of PGC 1126831 (excluded) with J=0.15, s = 0.00121 and J/s = 123, SBSG 1646+499 (excluded) with J=2.30, s = 0.00751 and J/s = 305 and Mrk 841 (selected) with J=2.02, s = 0.000061 and J/s = 33487 in the $r$ band.}
\end{figure*}

The main science goal of ZTF is to discover young supernovae nightly and search for rare and exotic transients \citep{2019PASP..131a8003M}. ZTF has $g$, $r$, and $i$ filters, however for most targets, only $g$ and $r$ bands have adequate data for the broadband PRM. Therefore in this work, the redshift of AGNs is limited to be lower than 0.09 so that the ${\rm H}\alpha$ emission line can fall into the $r$ band. Figure 1 shows the spectrum of one AGN and the transmission functions of the ZTF broad bands. We use the $r$ band to trace the H$\alpha$ line and the $g$ band to trace the continuum\footnote{The $i$ band data from ZTF are usually inadequate for the PRM, so we do not use them to trace the continuum.}. To combine the ZTF public and private data \citep[ZTF Image Service;][]{https://doi.org/10.26131/irsa539} which have different cadences, the lightcurves are resampled into the mean values of one-day bins. At $z < 0.09$, most AGNs are extended sources in morphology, the seeing and airmass have significant impacts on the accuracy of the photometry. Only part of AGNs in ZTF have sufficiently accurate data for broadband PRM. So in this work, we use two criteria to select the targets for PRM from ZTF. To select AGNs with sufficient variability, we use the Welch-Stetson J Variability Index \citep{1993AJ....105.1813W} to examine the variability of AGNs. The J index is composed of the relative error ($\delta$) and a weighting factor ($w_i$). The relative error is defined by \citet{1996PASP..108..851S} as
\begin{equation}
\delta_i= \frac{f_i-\overline{f}}{\sigma_{f,i}}\sqrt{\frac{n}{n-1} }.
\end{equation}
Here $n$ is the number of observations, $\sigma_{f,i}$ is the measurement error and $\overline{f}$ is the mean flux of the lightcurve. To reduce the influence of a very large flux change within a few data points which is unphysical, the weighting factor is defined as
\begin{equation}
  w_i=\left[1+(\frac{\delta_i}{2})^2\right]^{-1}.
\end{equation}
The J index is defined as
\begin{equation}
  \mathrm{J}=\frac{\sum w_i \mathrm{sgn}(\delta_i^2-1) \sqrt{\left\lvert \delta_i^2-1 \right\rvert } }{\sum w_i }.
\end{equation}

Here sgn simply returns the sign of the value. $\rm J<0$ means that the variability is dominated by the uncertainties of the observation. Because ZTF data have different numbers of observations for different targets, and the J index is only related to the flux while not related to the observation epoch, some AGNs with a lot of observations and sufficient variability may have small J index. However, some AGNs with very unsmooth lightcurves may still have large J index (see the middle panel of Figure 2 as an example). Therefore, only using the J index to select targets is not sufficient for the ZTF data.

In addition to evaluating the variability with the J index, it is also necessary to select AGNs with smooth lightcurves to reduce the uncertainties in both the observations and photometric data processing as much as possible. After excluding the data points that deviate from the mean flux by $3\sigma$, we use the second derivative to evaluate the smoothness of lightcurves. The first order variation is
\begin{equation}
  \delta f_i=\frac{1}{\overline{f}}\frac{f_{i+1}-f_i}{t_{i+1}-t_i}.
\end{equation}
Here $t_i$ is the epoch of the observation. To avoid the influence of highly unevenly-sampled data due to the lack of observations such as seasonal gaps, the intervals of the points to calculate the second derivative are limited to $t_{i+2}-t_i<20$ days. The smoothness index $s$ is defined as
\begin{equation}
  s=\sum \left\lvert \frac{\delta f_{i+1}-\delta f_i}{t_{i+1}-t_i}\right\rvert \frac{\sqrt{ (f_{i+2}-f_{i+1})^2+(f_{i+1}-f_i)^2} } {\overline{f}}.
\end{equation}

\begin{figure}[!h]
  \includegraphics[width=1.0\linewidth]{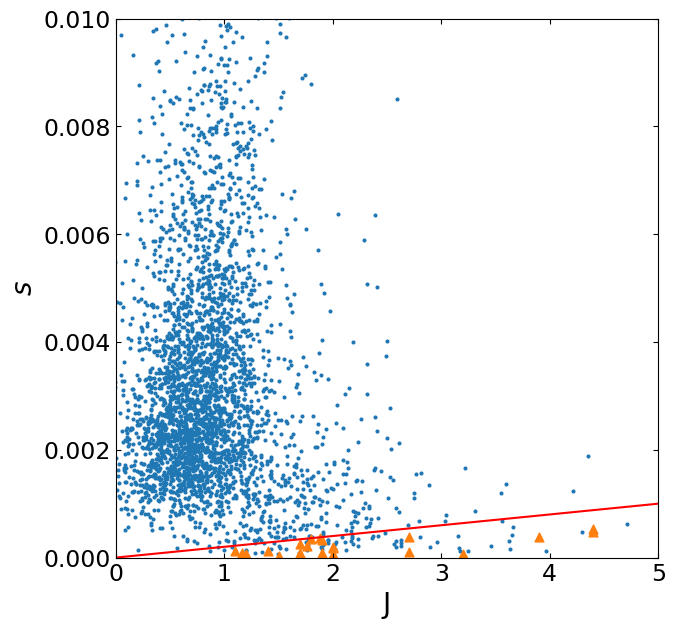}
  \caption{The distributions of ${\rm J}$ and $s$ parameters of 3537 ZTF AGNs. The orange triangles represent 23 selected AGNs for which we obtain the H$\alpha$ lags.}
\end{figure}

We exclude AGNs with large $s$ values, which have very large irregular variabilities during very short times. The AGN lightcurves are usually well modeled by the damped random walk \citep[DRW;][]{2009ApJ...698..895K,2010ApJ...708..927K,2010ApJ...721.1014M,2013ApJ...765..106Z}. However, the rapid, large, and irregular variabilities of AGNs are often not consistent with the DRW model \citep{2009ApJ...698..895K} and are probably caused by the errors in the observations for these ZTF targets. Considering the variabilities (J index) as well as the smoothness ($s$) of the lightcurves, we select AGNs with ${\rm J}>1$ and ${\rm J}/s>4000$. 
From 3537 AGNs which have more than 100 observational epochs in both the $g$ and $r$ bands, 123 AGNs are selected by these criteria. Figure 2 shows the example lightcurves of one selected AGN and two excluded AGNs.

One of the purposes of this work is to verify the ICCF-Cut approach we recently proposed in Paper I, therefore we select 23 AGNs that have previous SRM results \citep{2011MNRAS.416..225S, 2018ApJ...869..142D, 2019ApJ...886...42D, 2021ApJS..253...20H, 2021ApJ...918...50L, 2022ApJS..263...10L, 2022ApJS..262...14B, 2022ApJ...925...52U} for further ${\rm H}\alpha$ time lag calculation. Figure 3 presents our final targets.  The parameters of the selected 23 AGNs are shown in Table 1. 

In addition, our H$\alpha$ time lag measurement requires a single-epoch spectrum. Some of these AGNs have the spectra from Large Sky Area Multi-Object Fiber Spectroscopic Telescope \citep[LAMOST;][]{2015RAA....15.1095L}, the Sloan Digital Sky Survey \citep[SDSS;][]{2000AJ....120.1579Y} and the BAT AGN Spectroscopic Survey \citep[BASS;][] {2017ApJ...850...74K}. For those AGNs without publicly available spectra, we obtain their single epoch spectra using the Beijing Faint Object Spectrograph and Camera \citep[BFOSC;][]{2016PASP..128k5005F} of the Xinglong 2.16-m telescope, where we use the Grism 4 with a dispersion of $\SI{198}{\angstrom}/{\rm mm}$ and the slit width of $1\arcsec.8$. We also use the Yunnan Faint Object Spectrograph and Camera \citep[YFOSC;][]{2019RAA....19..149W} of the Lijiang 2.4m-telescope in China, where we use the Grism 3 with a dispersion of $\SI{215}{\angstrom}/{\rm mm}$ and the slit width of $2\arcsec.5$. The spectra are reduced by the standard IRAF \citep{1986SPIE..627..733T,1993ASPC...52..173T} routine. In particular, the data are reduced with PyFOSC \citep{yuming_fu_2020_3915021}, a pipeline toolbox based on PyRAF \citep{2012ascl.soft07011S} for the long-slit spectroscopy\footnote{Please refer to \href{https://github.com/rudolffu/pyfosc/tree/master/src} {https://github.com/rudolffu/pyfosc/tree/master/src} for detailed descriptions on the iraf tasks that have been used.}. The host galaxy contribution is about $30\%\sim 40\%$, but with large uncertainties. For some AGNs, we fail to decompose the host galaxy components due to the low spectral resolution of the spectra. All the spectra are presented in Appendix (Figure A1) and the epochs of spectroscopic data for each source are listed in Table 2-4.

\begin{deluxetable}{ccccc}[!h]
\centering
\tablecaption{The list of 23 selected AGNs. }
\tablehead{
\colhead{Name} & \colhead{Redshift} & \colhead{${\rm J}$} &\colhead{$s(10^5 \cdot {\rm s}^{-2})$} & \colhead{${\rm J}/s$}
}
\startdata
IRAS 04416+1215 & 0.089 & 3.19 & 6.36 & 50126\\
MCG +08-11-011 & 0.021 & 4.36 & 52.8 & 8254\\
Mrk 6 & 0.019 & 3.54 & 73.4 & 4822\\
Mrk 79 & 0.022 & 2.67 & 39.0  & 6910\\
Mrk 110 & 0.035 & 2.00 & 17.9 & 11137\\
Mrk 290 & 0.030 & 1.70 & 24.6 & 6916\\
Mrk 335 & 0.025 & 2.16 & 18 & 12018\\
Mrk 486 & 0.038 & 1.93 & 8.6 & 22511\\
Mrk 493 & 0.031 & 1.86 & 32.9 & 5627\\
Mrk 509 & 0.034 & 2.71 & 11.1 & 24448\\
Mrk 817 & 0.031 & 1.36 & 12.4 & 10966\\
Mrk 841 & 0.036 & 2.02 & 6.1 & 33487\\
Mrk 1044 & 0.016 & 1.98 & 28.4 & 6981\\
NGC 5548 & 0.017 & 2.12 & 32.6 & 6508\\
PG 0007+106 & 0.087 & 1.79 & 34.1 & 5255\\
PG 0844+349 & 0.064 & 1.49 & 3.5 & 42620\\
PG 1211+143 & 0.081 & 1.16 & 9.2 & 12643\\
PG 1229+204 & 0.064 & 1.20 & 7.3 & 16511\\
PG 1426+015 & 0.087 & 1.89 & 7.0 & 27136\\
PG 1440+356 & 0.077 & 2.88 & 10.1 & 28439\\
PG 2130+099 & 0.063 & 1.20 & 6.25 & 19215\\
PGC 3095715 & 0.017 & 4.69 & 39.8 & 11784\\
PGC 3096594 & 0.050 & 3.15 & 10.4 & 30288\\
\enddata
\end{deluxetable}
\vspace{-1.5cm}

\section{Time Lag Calculations}

For the broadband PRM, the continuum is dominant and the broad emission line  only contributes a small fraction of the total flux in the broadband. We need to remove the contribution of the continuum and extract the broad emission line from the broad band. In Paper I, we proposed the ICCF-Cut approach\footnote{The code for the ICCT-Cut process is provided at \href{https://github.com/PhotoRM/ICCF-Cut} {https://github.com/PhotoRM/ICCF-Cut}}, which is a combination of a cut procedure and the ICCF method \citep{1987ApJS...65....1G, 1994PASP..106..879W, 1998PASP..110..660P}. We extract the lightcurve of broad emission line in the broadband via the cut procedure, and then we use the ICCF method to compute the time lag of the broad emission line corresponding to the continuum. The whole process described above is named as the ICCF-Cut process in this paper. We assume that the continuum flux in the line band (in this work the ZTF r band) equals to a fixed fraction $\alpha$ of the flux in the continuum band (the ZTF g band) for each AGN. The extracted H$\alpha$ lightcurve can be expressed as 

\begin{equation}
  L_{{\rm H}\alpha}(t)=L_r(t)-\alpha L_g(t).
\end{equation}
Here $L_{\rm line}(t)$, $L_{\rm cont}(t)$, and $L_{{\rm H}\alpha}(t)$ are the lightcurves of the line band, continuum band, and ${\rm H}\alpha$ emission line respectively. In this equation, we ignored the contribution of other emission lines such as ${\rm H}\beta$ and [OIII] in the g band, since they are much weaker than the continuum and the broad ${\rm H}\alpha$ line. In Paper I we also conducted the simulations and found that the effect of other emission lines can be neglected. The fraction $\alpha$ value can be obtained from the single-epoch spectrum and the broad band lightcurves:
\begin{equation}
  \alpha=\left(1-\frac{F_{{\rm H}\alpha}}{F_{\rm line}}\right)\min\left(\frac{L_{{\rm line},t}}{L_{{\rm cont},t}}\right).
\end{equation}
Here $F_{\rm cont}$, $F_{\rm line}$ and $F_{{\rm H}\alpha}$ are the fluxes of continuum band, line band and ${{\rm H}\alpha}$ line obtained from the integral of the single-epoch spectrum, and $L_{{\rm line},t}$ and $L_{{\rm cont},t}$ are the fluxes of line band and continuum band at each data point obtained from the linearly interpolated photometric lightcurves. 
For the 23 ZTF AGNs, some of their spectra and photometry are not synchronous. Besides, the observation strategies such as the slit width and aperture size may cause the differences between the spectroscopy and photometry. So the latter term of Equation 7 is added to adjust the ratio calculated from the spectra (i.e., $1-F_{{\rm H}\alpha}/F_{\rm line}$) to the photometry data, and the minimum function is used to make sure that the extracted ${{\rm H}\alpha}$ lightcurve contains all the contributions of the ${{\rm H}\alpha}$ emission line (see detailed discussion in Paper I).

Same as in Paper I, we also consider the influence of the host galaxy and season gaps. We use the flux variation gradient \citep[FVG;][]{1981AcA....31..293C, 1992MNRAS.257..659W, 2012A&A...545A..84P} to determine the contribution of the host galaxy. Because the observational duration of ZTF is as long as several years, there are large season gaps without observations. The spectral index of the continuum and the value of $\alpha$ may change during the long observation durations. We divide the lightcurve into several parts according to the season gaps. For each part, the value of $\alpha$ is adjusted according to the ratio between line band and continuum band fluxes.

\begin{figure*}[!htb]
  \includegraphics[width=1.0\linewidth]{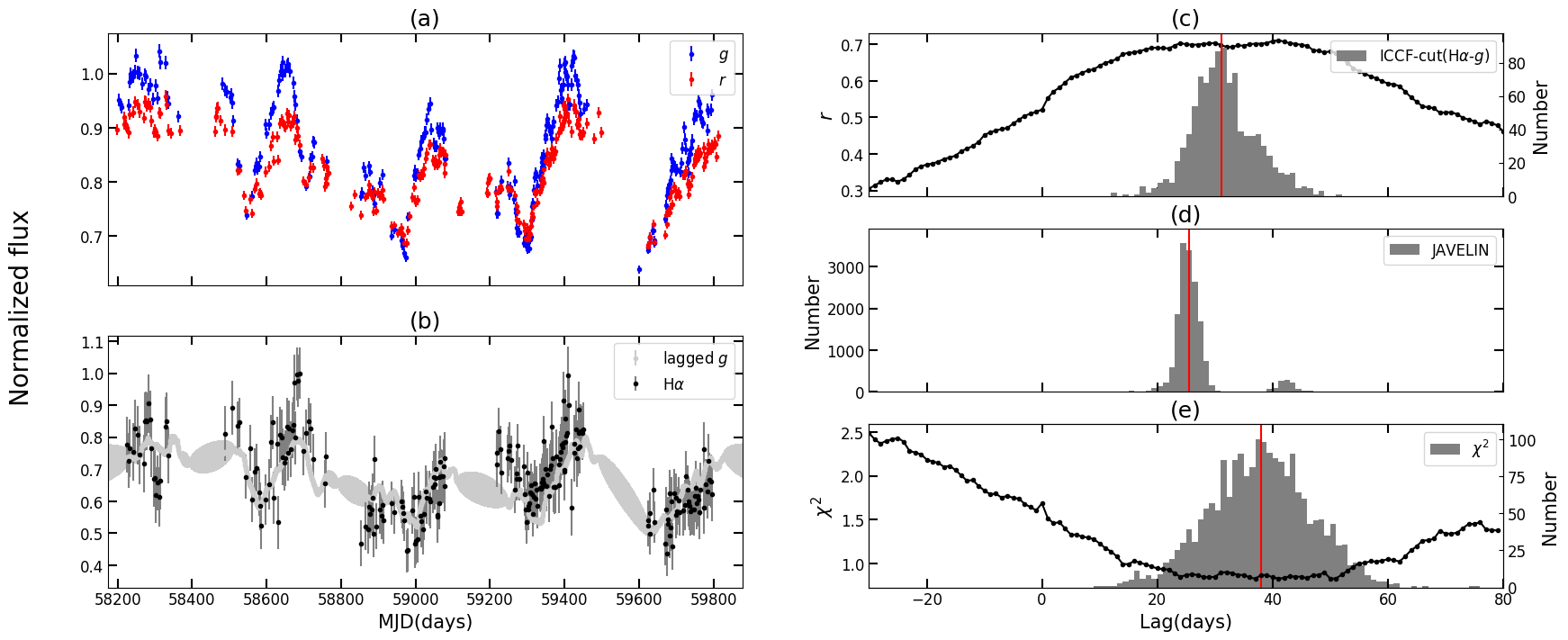}
  \caption{The lightcurves and lag distributions for Mrk 841. The upper left panel shows the lightcurves of the continuum band ($g$) and line band ($r$). The bottom left panel shows the extracted H$\alpha$ lightcurve, compared with the lagged continuum band lightcurve. The right three panels show the lag distributions between the continuum band and extracted H$\alpha$ lightcurves with the ICCF-Cut, JAVELIN and $\chi^2$ methods respectively. The red line represents the median value of the lag distribution.}
\end{figure*}

\begin{figure*}[htb!]
  \includegraphics[width=1.0\linewidth]{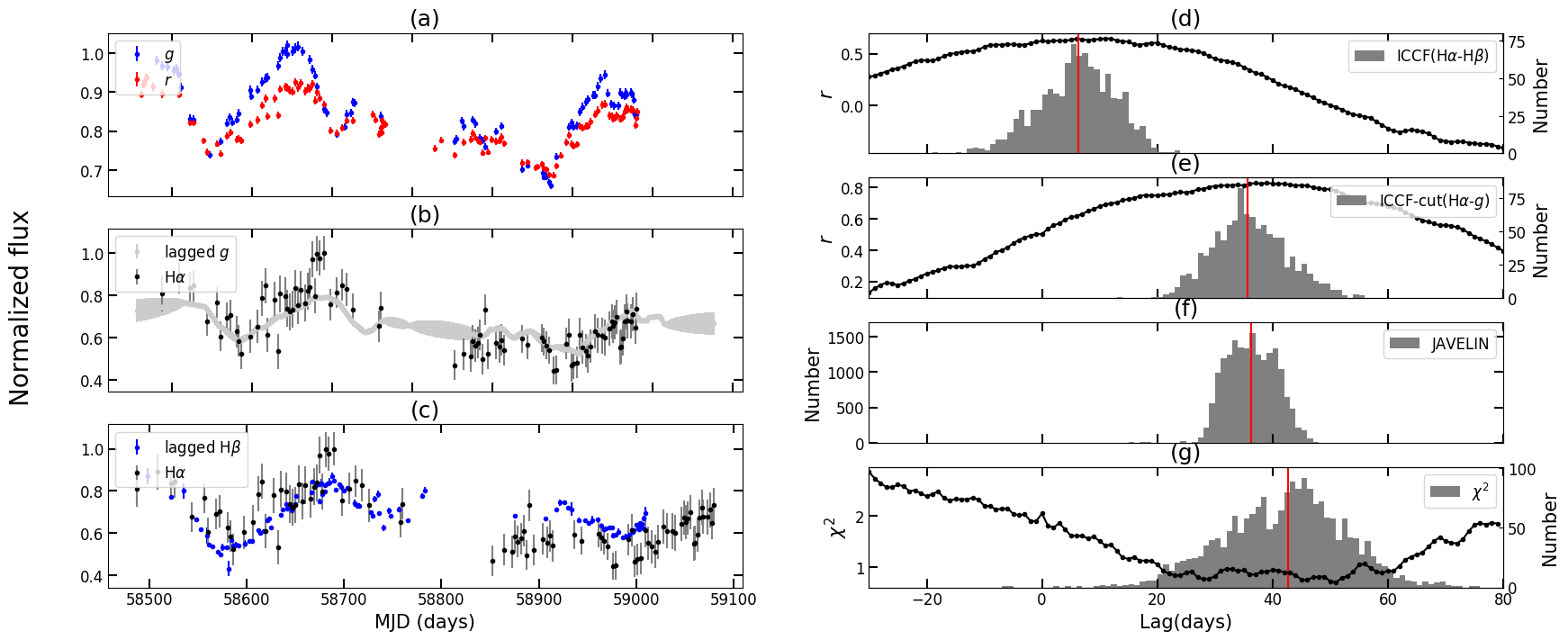}
  \caption{The lightcurves and lag distributions for Mrk 841 in the period with simultaneous SRM observation. The upper left panel shows the lightcurves of the continuum band ($g$) and line band ($r$). The middle and bottom left panels show the extracted H$\alpha$ lightcurves compared with the lagged continuum band and SRM H$\beta$ broad line lightcurves. The upper right panel shows the lag distribution between the SRM H$\beta$  and extracted H$\alpha$ lightcurves given by ICCF. The rest three panels show the lag distributions between the continuum band and extracted H$\alpha$ lightcurves with the ICCF-Cut, JAVELIN and $\chi^2$ methods respectively. The red line represents the median value of the lag distribution. }
\end{figure*}

\begin{figure*}[!htb]
  \includegraphics[width=1.0\linewidth]{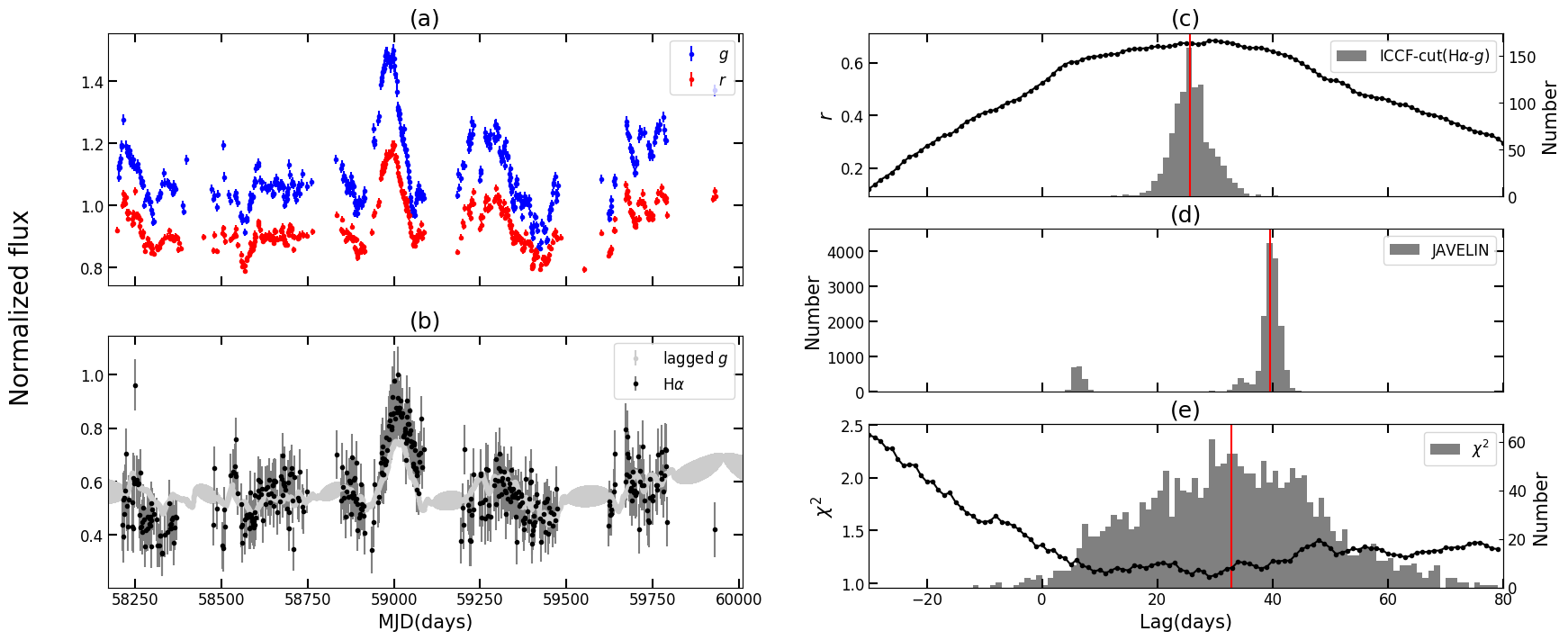}
  \caption{Same as Figure 4 but for Mrk 817.}
\end{figure*}

\begin{figure*}[!htb]
  \includegraphics[width=1.0\linewidth]{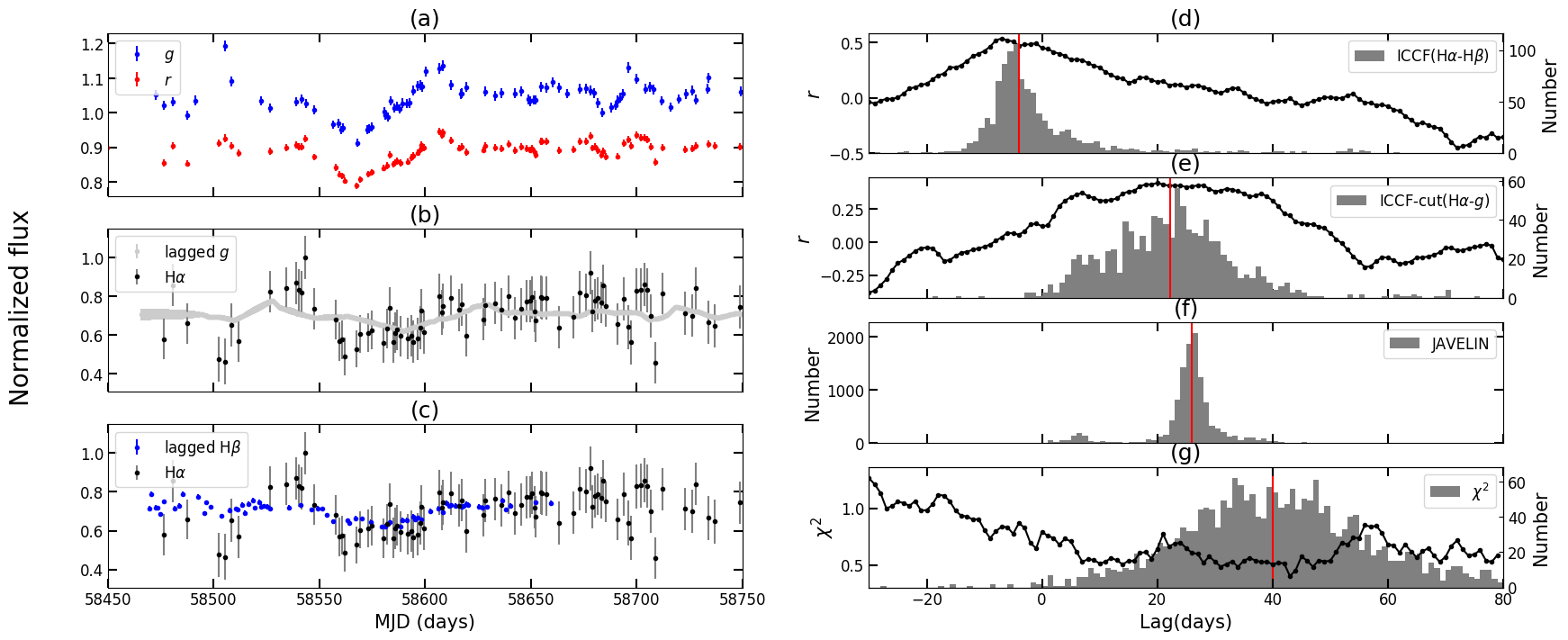}
  \caption{Same as Figure 5 but for Mrk 817.}
\end{figure*}

\begin{figure*}[!htb]
  \includegraphics[width=1.0\linewidth]{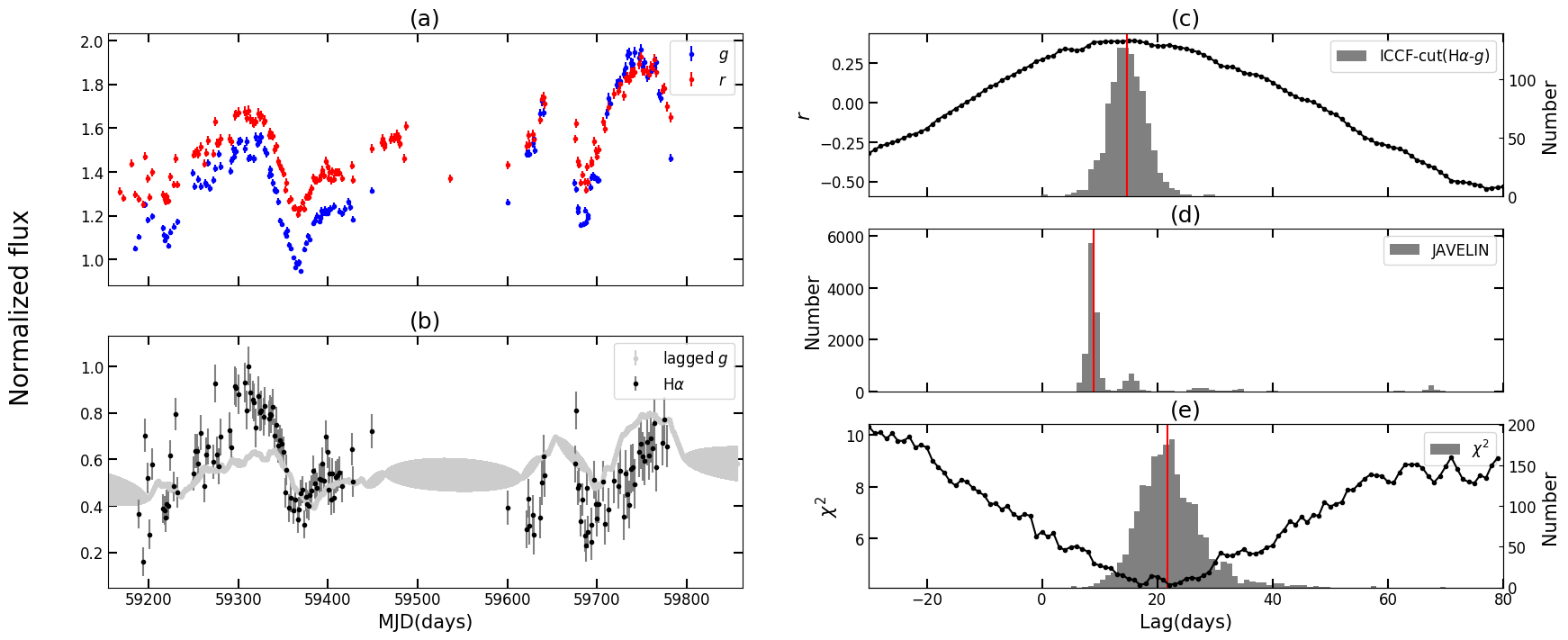}
  \caption{Same as Figure 4 but for NGC 5548.}
\end{figure*}

\begin{figure*}[!htb]
  \includegraphics[width=1.0\linewidth]{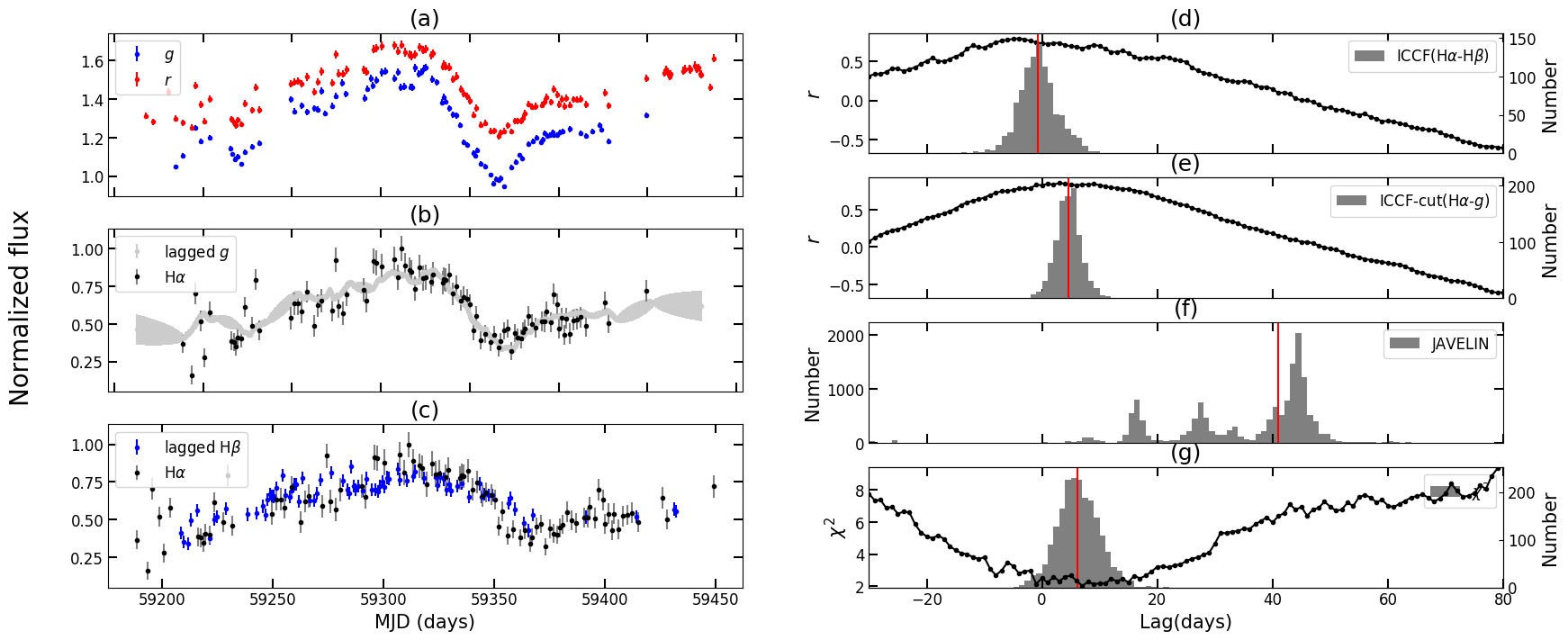}
  \caption{Same as Figure 5 but for NGC 5548, except that the SRM lightcurve for comparison here is H$\alpha$ broad line lightcurve.}
\end{figure*}

\begin{figure*}[htb!]
  \includegraphics[width=1.0\linewidth]{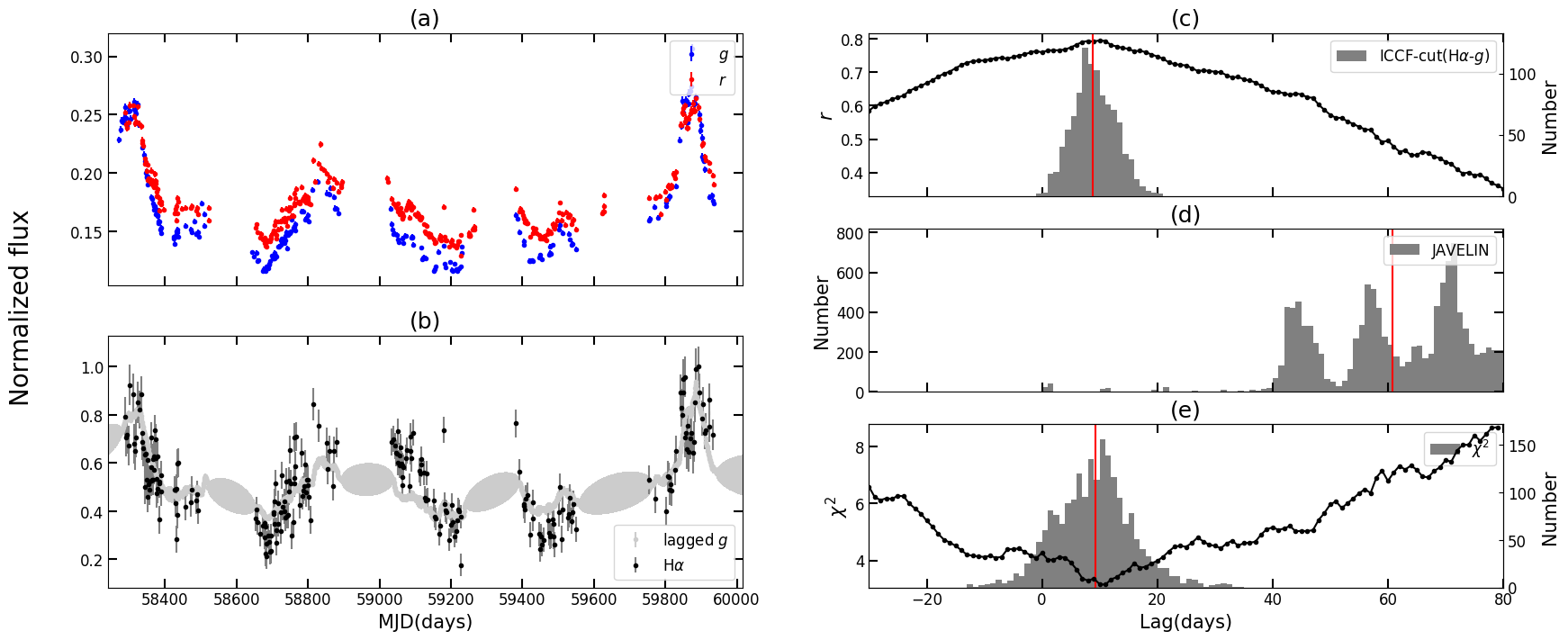}
  \caption{Same as Figure 4 but for PG 0007+106.}
\end{figure*}

\begin{figure*}[htb!]
  \includegraphics[width=1.0\linewidth]{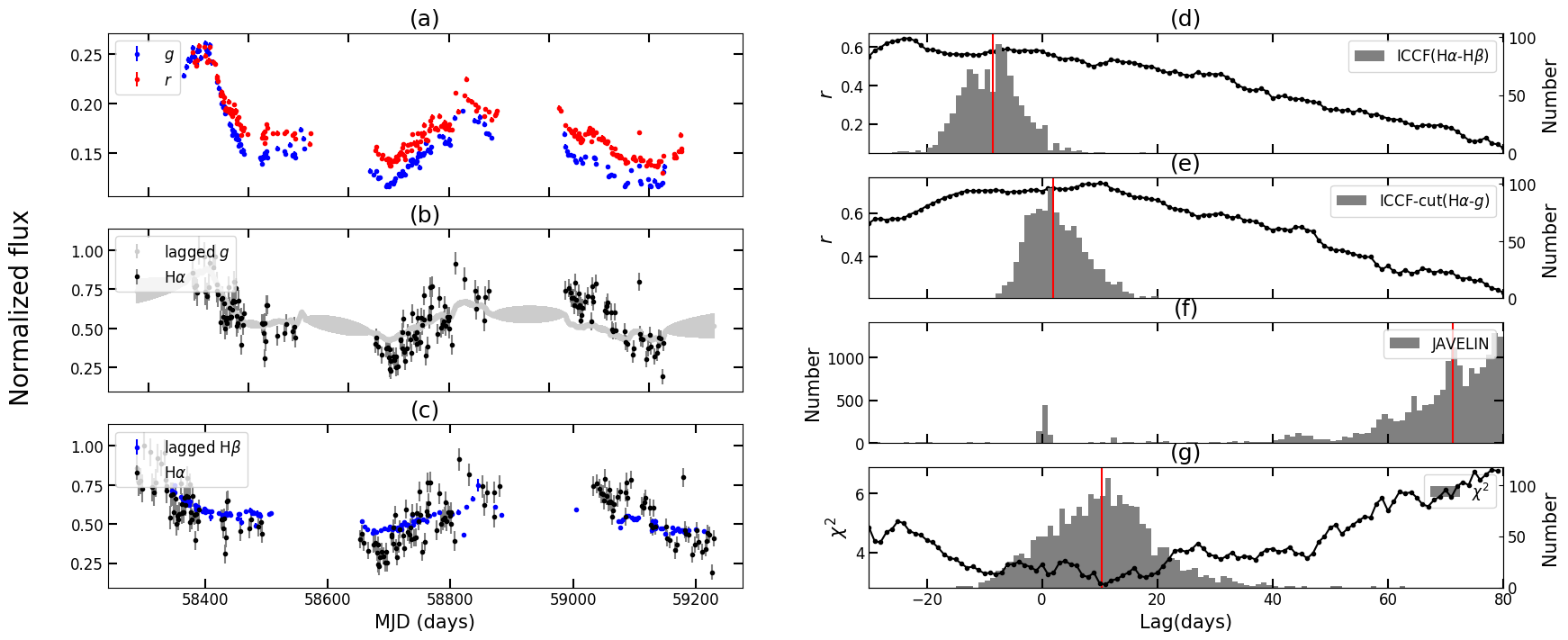}
  \caption{Same as Figure 5 but for PG 0007+106.}
\end{figure*}

\begin{figure*}[htb!]
  \includegraphics[width=1.0\linewidth]{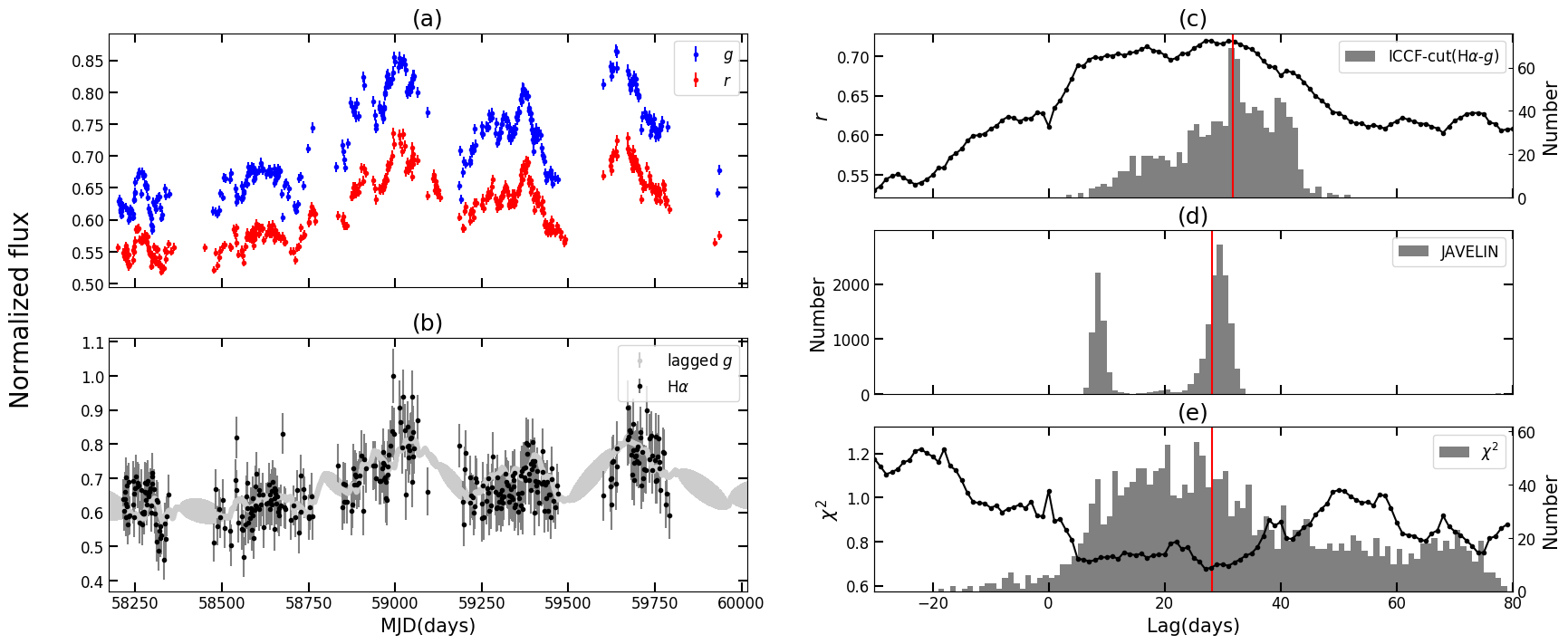}
  \caption{Same as Figure 4 but for PG 1440+356.}
\end{figure*}

\begin{figure*}[htb!]
  \includegraphics[width=1.0\linewidth]{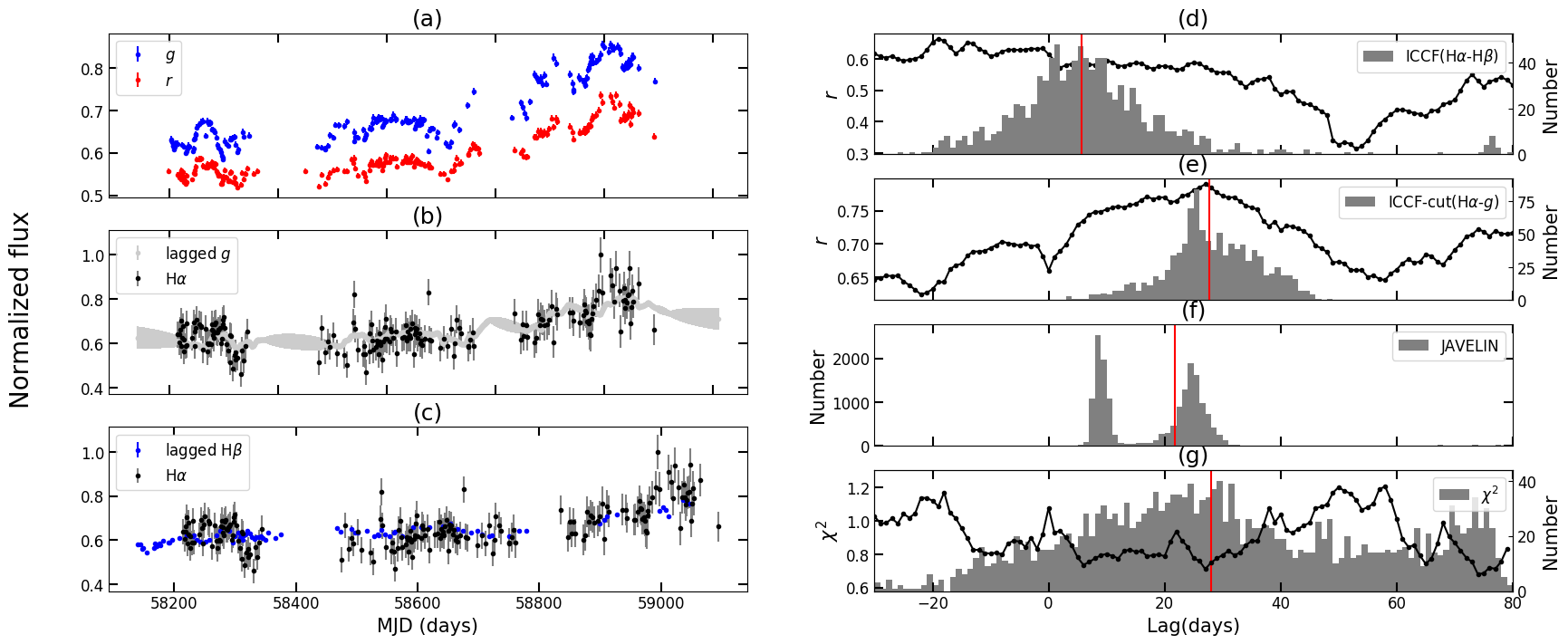}
  \caption{Same as Figure 5 but for PG 1440+356.}
\end{figure*}

Besides the ICCF-Cut process, we also used the Just Another Vehicle for Estimating Lags In Nuclei \citep[JAVELIN:][]{2011ApJ...735...80Z,2013ApJ...765..106Z,2016ApJ...819..122Z} and $\chi^2$ methods to calculate the H$\alpha$ time lags of the AGNs. JAVELIN uses the damped random walk (DRW) process to model the AGN lightcurves and uses the Markov Chain Monte Carlo (MCMC) DRW processes to reproduce the lightcurves and obtain the distributions of the parameters including the time lag. We use the JAVELIN Pmap Model \citep{2016ApJ...819..122Z} which can separate the continuum component and the strong emission line component blended in line band via thousands of fitting. Since there are a lot of fitting parameters, the results of JAVELIN are more dependent on the quality of lightcurves and may have large scatters for lightcurves with short durations and/or large errors. Because the ZTF is the photometric sky survey with a 48 inch telescope, the photometric accuracy should be considered carefully. 
The $\chi ^2$ method \citep{2013A&A...556A..97C,2022ApJS..262...14B} calculates the correlation between two lightcurves with weighting the observational uncertainties. It can be used to evaluate the influence of errors and the feasibility of the broadband PRM with large errors. More details on the ICCF-Cut, JAVELIN and $\chi^2$ methods can be found in Paper I.

Combining the ICCF-Cut, JAVELIN and $\chi ^2$ methods, we obtain the H$\alpha$ time lags of 23 AGNs which all have SRM results. If the 3 methods can give consistent results of H$\alpha$ lags, we consider the PRM results are more reliable. However, since the ICCF-Cut process and the $\chi^2$ method share the similar principle and use the same pair of lightcurves (i.e., continuum band lightcurve and the extracted H$\alpha$ lightcurve) to calculate the time lags, it's expected that the 2 methods can give consistent results in most cases. However, there are also some cases when the results given by the 2 methods are not consistent, which are mainly due to the large errors of ZTF lightcurves. In such cases, the $\chi^2$ method may not be as suitable as the ICCF-Cut process.

With these considerations, we divide these AGNs into three lists. List I has 5 AGNs which all have simultaneous SRM observation data. For these AGNs, we can compare their simultaneous Balmer line lags and the shape of emission line lightcurves with SRM. It can increase the reliabilities of our H$\alpha$ lag results furthermore. For AGNs in List II, the lag distributions of the ICCF-Cut, JAVELIN and $\chi ^2$ methods or at least those of the ICCF-Cut and JAVELIN methods are consistent, and the extracted H$\alpha$ lightcurves are similar to the lagged $g$ band continuum lightcurves (here the lagged lightcurve refers to the lightcurve that is lagged as well as scaled). Meanwhile, the obtained broadband PRM H$\alpha$ lags are the same as or slightly larger than the SRM H$\beta$ lags, which is consistent with the structure of BLR \citep{2004ApJ...606..749K} and the results of previous works \citep{2000ApJ...533..631K,2010ApJ...716..993B,2012ApJ...755...60G}. For AGNs in List III, the JAVELIN results for some AGNs are not consistent with those from any of other 2 methods. Some AGNs have three consistent lag distributions, but the obtained H$\alpha$ lags are much larger than SRM H$\beta$ lags. These deviations mean that the results of AGNs in List III are not very convincing. 

For Mrk 841 in List I, we first calculate the time lag distributions with the ICCF-Cut, JAVELIN and $\chi ^2$ methods for the whole observational duration. The obtained H$\alpha$ lag distributions from three methods are consistent. The extracted H$\alpha$ lightcurve is similar to the lagged continuum $g$ band lightcurve, as shown in Figure 4 panel (b). Then we calculate the time lag distribution for part of the lightcurves which have simultaneous SRM H$\beta$ observations \citep{2022ApJS..262...14B}. By comparing with the SRM H$\beta$ lightcurve, we can find that the shape of the extracted H$\alpha$ lightcurve is very similar to the SRM H$\beta$ lightcurve, and the ICCF result between the extracted H$\alpha$ and SRM H$\beta$ lightcurves is close to zero (see panels (c) and (d) in Figure 5). It proves the reliability of the extracted H$\alpha$ lightcurve. The time lags given by three methods of this part of lightcurves are consistent with each other and close to the lag results of the whole lightcurves, which further assures the reliability of the obtained H$\alpha$ time lags for Mrk 841. 

For Mrk 817, the results are shown in Figure 6 and Figure 7. The JAVELIN lag of the whole lightcurves is slightly larger than others and the $\chi^2$ lag distributions have larger scatters compared to other methods. But the similar shape of the extracted H$\alpha$, lagged $g$ and lagged SRM H$\beta$ lightcurves \citep{2021ApJ...918...50L} for both the whole and part of lightcurves, as well as the consistent lag distributions from at least two methods, can indicate the reliability of the H$\alpha$ time lags for Mrk 817.

For NGC 5548, there are no enough data for some observational seasons. We only use two years of data which have enough observations. For the whole lightcurves, three methods also give consistent results, though the extracted H$\alpha$ lightcurve doesn't match very well with the lagged g band lightcurve. For part of the lightcurves which have simultaneous SRM H$\alpha$ observations \citep{2022ApJS..263...10L}, the JAVELIN result is not consistent with the other 2 results, which means that the results for part of lightcurves are not as reliable as those for the whole lightcurves. The reason may be that part of the lightcurves have less data points and shorter duration.

For PG 0007+106, the qualities of lightcurves are not as good as other four AGNs. From Table 1, the value of J/s for PG 0007+106 is the smallest among five AGNs in List I, which may account for the inconsistent JAVELIN results with other two methods and the large scatters of JAVELIN lag distributions for both the whole lightcurves (Figure 10) and part of the lightcurves (Figure 11, the simultaneous H$\beta$ lightcurve is from \citet{2022ApJS..262...14B}). Therefore, we consider the results of PG 0007+106 not as reliable as other AGNs, which is also confirmed by the comparison between PRM H$\alpha$ lags and SRM H$\beta$ lags (see Table 2).

\begin{deluxetable*}{cccccccccc}[htb!]
\vspace{-1cm} 
\centering
\tablecaption{The ${\rm H}\alpha$ lag results (in days) of 5 AGNs in List I (with simultaneous SRM lags).}
\tablehead{
\colhead{Name} & \colhead{log $L_{5100}$(erg/s)} & \colhead{${\rm H}\alpha$ ratio} & \colhead{Spec Epoch} & \colhead{$ICCF-Cut$} & \colhead{$\chi^2$} & \colhead{JAVELIN} & \colhead{ SRM lag}
}
\startdata
Mrk 817 & $43.82\pm 0.12$ & 0.12 & 59639$^d$ & $25.7_{-2.7}^{+3.5}$ & $32.8_{-16.0}^{+15.2}$ & $39.6_{-4.6}^{+1.4}$\\
 & & & & $22.2_{-11.1}^{+9.4}$ & $40.0_{-14.9}^{+15.7}$ & $26.0_{-2.6}^{+2.5}$ & $19.9_{-6.7}^{+9.9}\ ^{(1)}$\\
Mrk 841 & $43.88\pm 0.10$ & 0.18 & 54269$^c$ & $31.2_{-4.6}^{+6.1}$ & $38.0_{-8.8}^{+8.5}$ & $25.5_{-1.6}^{+2.1}$\\
 & & & & $35.6_{-5.8}^{+7.1}$ & $42.8_{-11.6}^{+9.6}$ & $36.3_{-4.5}^{+4.5}$ & $26.0_{-2.2}^{+2.0}\ ^{(2)}$\\
NGC 5548 & $43.35\pm 0.14$ & 0.14 & 53859$^a$ & $14.8_{-3.0}^{+3.1}$ & $21.8_{-4.6}^{+5.3}$ & $9.0_{-0.8}^{+7.0}$\\
 & & & & $4.6_{-2.0}^{+2.0}$ & $6.2_{-3.3}^{+3.6}$ & $41.0_{-23.0}^{+4.2}$ & $14.5_{-5.8}^{+7.1}\ ^{(3)}$\\
PG 0007+106 & $43.21\pm 0.15$ & 0.21 & 56156$^c$ & $8.9_{-4.0}^{+3.9}$ & $13.5_{-6.1}^{+6.8}$ & $60.8_{-15.5}^{+11.6}$\\
& & & & $1.9_{-4.3}^{+5.4}$ & $10.3_{-9.4}^{+8.3}$ & $71.3_{-12.1}^{+6.5}$ & $30.9_{-2.4}^{+2.5}\ ^{(2)}$\\
PG 1440+356 & $44.73\pm 0.21$ & 0.19 & 60084$^e$ & $30.8_{-12.5}^{+8.1}$ & $28.5_{-17.7}^{+30.6}$ & $28.2_{-19.6}^{+2.3}$\\
& & & & $27.7_{-6.4}^{+8.6}$ & $28.0_{-23.0}^{+35.1}$ & $21.8_{-13.3}^{+4.1}$ & $34.5_{-12.4}^{+10.8}\ ^{(4)}$ \\
\enddata
\tablecomments{For each AGN, the data in the second row represents the lag distributions for the part of lightcurves with simultaneous SRM observations. For NGC 5548, the SRM lag is the SRM H$\alpha$ lag. For other AGNs, the SRM lag is the SRM H$\beta$ lag. The references of the SRM lags: (1) \citet{2019ApJ...886...42D}, (2) \citet{2022ApJS..262...14B} , (3) \citet{2022ApJS..263...10L}, (4) \cite{2021ApJS..253...20H}, (5) \citet{2018ApJ...869..142D}, (6) \citet{2022ApJ...925...52U}, (7) \citet{2011MNRAS.416..225S}. The Spec Epoch column refers to the MJD time when the spectrum we used in this paper was taken, where the superscripts refer to different spectra sources: $^a$SDSS \citep{2000AJ....120.1579Y}; $^b$LAMOST \citep{2015RAA....15.1095L}; $^c$BASS \citep{2017ApJ...850...74K}; $^d$BFOSC \citep{2016PASP..128k5005F}; $^e$YFOSC \citep{2019RAA....19..149W}.}
\end{deluxetable*}

\begin{deluxetable*}{cccccccc}[htb!]
\vspace{-1cm} 
\centering
\tablecaption{The ${\rm H}\alpha$ lag results (in days) of 8 AGNs in List II.}
\tablehead{
\colhead{Name} & \colhead{log $L_{5100}$(erg/s)} & \colhead{${\rm H}\alpha$ ratio} & \colhead{Spec Epoch} & \colhead{$ICCF-Cut$} & \colhead{$\chi^2$} & \colhead{JAVELIN}  & \colhead{${\rm H}\beta$ lag}
}
\startdata
MCG +08-11-011 & $43.33\pm 0.11$ & 0.24 & 59286$^e$ & $32.9_{-6.5}^{+4.9}$ & $36.4_{-5.8}^{+6.2}$ & $35.3_{-0.9}^{+23.4}$ & $15.7_{-0.5}^{+0.5}\ ^{(1)}$  \\
Mrk 6 & $42.96\pm 0.19$ & 0.27 & 59639$^d$ & $25.6_{-0.4}^{+0.5}$ & $35.0_{-6.9}^{+6.4}$ & $43.4_{-10.8}^{+28.4}$ & $18.5_{-2.4}^{+2.5}\ ^{(5)}$  \\
Mrk 79 & $43.67\pm 0.11$ & 0.21 & 53733$^c$ & $19.1_{-6.4}^{+5.4}$ & $20.2_{-3.5}^{+3.8}$ & $17.3_{-2.9}^{+42.3}$ & $15.6_{-4.9}^{+5.1}\ ^{(1)}$  \\
Mrk 493 & $43.08\pm 0.12$ & 0.23 & 53141$^a$ & $15.0_{-4.7}^{+7.3}$ & $19.1_{-11.3}^{+9.7}$ & $16.6_{-8.0}^{+25.7}$ & $11.6_{-2.6}^{+1.2}\ ^{(1)}$  \\
Mrk 509 & $44.10\pm 0.09$ & 0.19 & 55324$^c$ & $77.5_{-14.2}^{+5.9}$ & $57.7_{-51.1}^{+21.3}$ & $61.7_{-27.0}^{+5.2}$ & $79.6_{-5.4}^{+6.1}\ ^{(1)}$  \\
Mrk 1044 & $43.12\pm 0.08$ & 0.17 & 55566$^c$ & $13.1_{-4.7}^{+16.0}$ & $61.4_{-35.0}^{+12.5}$ & $10.9_{-0.8}^{+30.5}$ & $10.5_{-2.7}^{+3.3}\ ^{(1)}$  \\
PG 0844+349 & $44.15\pm 0.21$ & 0.22 & 59970$^b$ & $20.7_{-4.9}^{+4.4}$ & $17.6_{-5.4}^{+5.8}$ & $21.3_{-1.4}^{+2.3}$  & $32.3_{-13.4}^{+13.7}\ ^{(1)}$ \\
PGC 3096594 & $43.68\pm 0.11$ & 0.22 & 54998$^c$ & $23.5_{-7.8}^{+7.7}$ & $55.2_{-36.1}^{+18.8}$ & $14.4_{-9.8}^{+2.3}$  & $14.4_{-1.9}^{+1.6}\ ^{(6)}$ \\
\enddata
\tablecomments{The meaning of the superscripts in the H$\beta$ lag column and the Spec Epoch column is the same as in Table 2.}
\end{deluxetable*}

\begin{deluxetable*}{cccccccc}[htb!]
\vspace{-1cm} 
\centering
\tablecaption{The ${\rm H}\alpha$ lag results (in days) of 10 AGNs in List III.}
\tablehead{
\colhead{Name} & \colhead{log $L_{5100}$(erg/s)} & \colhead{${\rm H}\alpha$ ratio} & \colhead{Spec Epoch} & \colhead{$ICCF-Cut$} & \colhead{$\chi^2$} & \colhead{JAVELIN}  & \colhead{${\rm H}\beta$ lag}
  }
\startdata
IRAS 04416+1215 & $43.60\pm 0.21$ & 0.18 & 54086$^a$ & $20.6_{-11.0}^{+4.9}$ & $19.7_{-26.9}^{+25.6}$ & $5.8_{-3.1}^{+32.8}$  & $13.3_{-1.4}^{+13.9}\ ^{(1)}$ \\
Mrk 110 & $43.55\pm 0.24$ & 0.32 & 52252$^a$ & $21.4_{-3.4}^{+2.7}$ & $19.8_{-4.1}^{+3.7}$ & $37.6_{-0.6}^{+1.7}$ & $25.6_{-7.2}^{+8.9}\ ^{(1)}$  \\
Mrk 290 & $43.48\pm 0.15$ & 0.13 & 52345$^a$ & $29.0_{-2.9}^{+3.6}$ & $29.1_{-8.9}^{+9.9}$ & $23.1_{-6.7}^{+2.8}$ & $8.7_{-1.0}^{+1.2}\ ^{(1)}$ \\
Mrk 335 & $43.76\pm 0.15$ & 0.21 & 55565$^c$ & $34.7_{-8.5}^{+5.7}$ & $38.3_{-14.7}^{+24.3}$ & $77.2_{-1.0}^{+0.9}$  & $14.0_{-3.4}^{+4.6}\ ^{(1)}$ \\
Mrk 486 & $43.85\pm 0.07$ & 0.20 & 56095$^b$ & $24.9_{-0.7}^{+0.7}$ & $19.1_{-8.4}^{+10.4}$ & $15.9_{-6.5}^{+18.6}$ & $23.7_{-2.7}^{+7.5}\ ^{(1)}$\\
PG 1211+143 & $44.73\pm 0.18$ & 0.27 & 60000$^e$ & $40.5_{-12.0}^{+10.4}$ & $35.5_{-19.5}^{+28.0}$ & $86.3_{-26.7}^{+9.0}$ & $33.0_{-5.5}^{+5.6}\ ^{(2)}$ \\
PG 1229+204 & $43.72\pm 0.05$ & 0.18 & 54481$^a$ & $33.3_{-9.4}^{+6.3}$ & $27.8_{-12.5}^{+16.6}$ & $21.3_{-8.9}^{+18.8}$  & $37.8_{-15.3}^{+27.6}\ ^{(1)}$ \\
PG 1426+015 & $43.63\pm 0.11$ & 0.19 & 55698$^c$ & $23.7_{-7.1}^{+1.0}$ & $40.8_{-26.9}^{+25.0}$ & $68.6_{-5.2}^{+4.5}$  & $95.0_{-37.1}^{+29.9}\ ^{(1)}$ \\
PG 2130+099 & $43.32\pm 0.12$ & 0.22 & 59915$^b$ & $41.7_{-10.4}^{+9.3}$ & $47.6_{-24.8}^{+20.7}$ & $75.6_{-65.4}^{+2.5}$ & $22.6_{-3.6}^{+2.7}\ ^{(1)}$  \\
PGC 3095715& $42.80\pm 0.15$ & 0.19 & 59245$^d$ & $24.5_{-5.3}^{+0.9}$ & $20.1_{-7.2}^{+10.2}$ & $12.8_{-1.3}^{+0.8}$  & $3.0_{-1.8}^{+0.4}\ ^{(7)}$ \\
\enddata
\tablecomments{The meaning of the superscripts in the H$\beta$ lag column and the Spec Epoch column is the same as in Table 2.}
\end{deluxetable*}
\vspace{-2.5cm}

For PG 1440+356, three methods give consistent results despite the large scatters of lag distributions. But by comparing our H$\alpha$ lags with simultaneous SRM H$\beta$ \citep{2021ApJS..253...20H}, the high correlation coefficient $r$ of ICCF-Cut in Figure 13 panel (e) as well as the close-to-zero ICCF lag in panel (d) still make the results reliable. The large scatters of the JAVELIN and $\chi^2$ methods probably mean that the ICCF-Cut process is more suitable for the broadband PRM. The obtained H$\alpha$ lags of 5 AGNs in List I are listed in Table 2.

For AGNs in List II, the extracted H$\alpha$ lightcurves are similar to the lagged $g$ band continuum and the lag distributions of the ICCF-Cut and JAVELIN are consistent. For 6 out of 8 AGNs in List II, the results of $\chi^2$ method are also consistent with the results of other two methods. For Mrk 1044 and PGC 3096594, the $\chi^2$ results are much larger than the ICCF-Cut and JAVELIN results, which may result from the large observational uncertainties of ZTF lightcurves. By comparing our broadband PRM results with non-simultaneous SRM results, it can be found that our obtained H$\alpha$ lags are the same as or slightly larger than the SRM H$\beta$ lags, which is consistent with previous works and theoretical predictions that the BLR size of the ${\rm H}\alpha$ line is usually larger than that of the ${\rm H}\beta$ emission line. The obtained H$\alpha$ lags of 8 AGNs in List II are listed in Table 3. Their lightcurves and lag distributions are shown in Appendix (Figure B1-B8).

For 8 out of 10 AGNs in List III, the results given by three methods are not well consistent. The $\chi^2$ results are consistent with the ICCF-Cut results, while the scatters of $\chi^2$ lag distributions are larger. However, the JAVELIN lag distributions show different peaks from ICCF-Cut and $\chi^2$ lag distributions. Therefore we consider these results not reliable as the results in List I and List II. For Mrk 335 and PGC 3095715, although the lag distributions of three methods are consistent with each other, their H$\alpha$ lags are more than 3 times than the previous SRM H$\beta$ lags, which are much larger than theoretical predictions. A reason may be that the broad band PRM results for these two targets may have large uncertainties. Another reason may be that the mean luminosities of the photometric observation periods are larger than those of the spectroscopic observation periods. According to the $R-L$ relation, the time lags in these photometric observation periods are larger than the time lags in the SRM observation periods. Besides, we note that for PGC 3095715, the H$\beta$ time lag from SRM is even smaller than the average cadence of ZTF lightcurves, so it is reasonable that we fail to obtain such small lags via ZTF lightcurves. The obtained H$\alpha$ lags of 10 AGNs in List III are listed in Table 4. Their lightcurves and lag distributions are shown in Appendix (Figure C1-C10).

\section{Discussion}

\begin{figure}
  \includegraphics[width=1.0\linewidth]{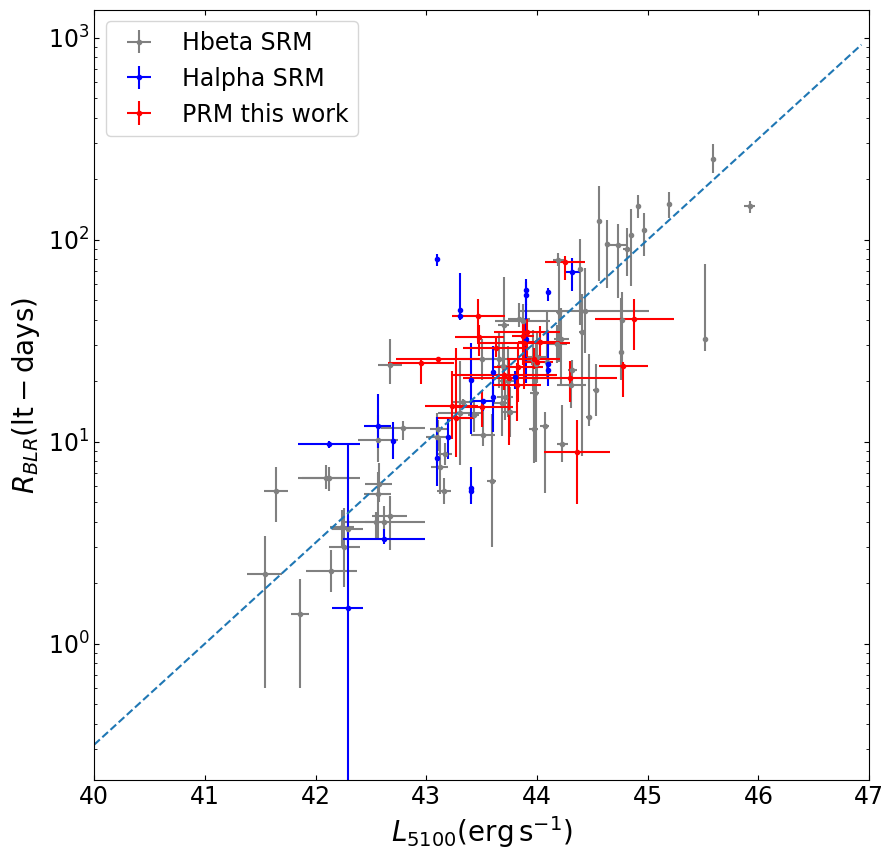}
  \caption{The $R_{BLR}-L_{5100}$ relationship of the AGNs from the broadband PRM (red points from this work), SRM for ${\rm H}\alpha$ line (blue points) and SRM for ${\rm H}\beta$ line (black points) \citep{2019ApJ...886...42D}. The ZTF ${\rm H}\alpha$ PRM time lags are obtained with the ICCF-Cut process. The dash line is the R-L relation given by \citet{2019FrASS...6...75P}.}
  \end{figure}
  
\begin{figure*}
  \includegraphics[width=1.0\linewidth]{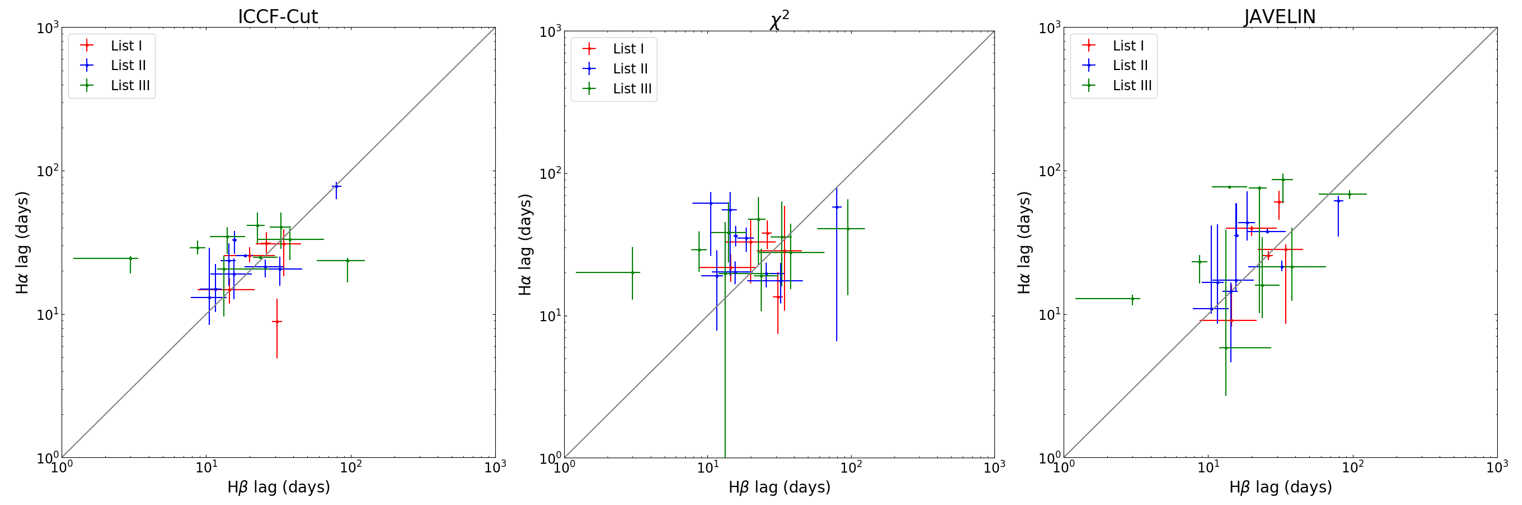}
  \caption{The comparisons of the ${\rm H}\alpha$ and ${\rm H}\beta$ time lags for 23 AGNs with the SRM results. The red, blue and green points represent the AGNs in List I,  List II, and List III respectively. From left to right, the obtained H$\alpha$ lags are from the ICCF-Cut, $\chi^2$ and JAVELIN methods. The solid line represents the one-to-one ratio.}
\end{figure*}

After correcting the extinction of the Milky Way \citep{1998ApJ...500..525S}, we use the mean flux values of the $g$ and $r$ bands and the spectra to obtain the continuum luminosity $L_{5100}$. The uncertainty of the continuum luminosity is calculated from the propagation of the flux errors, which are set as the variability in the $g$ and $r$ bands. We then plot the distribution of $R_{\rm{BLR}}$ for the 23 AGNs versus their $L_{5100}$ values. The result is shown in Figure 14, where we also list some results from previous SRM works \citep{2000ApJ...533..631K,2010ApJ...716..993B,2017ApJ...849..146G}. Our results from the ${\rm H}\alpha$ PRM are mostly consistent with the commonly adopted $R_{\rm{BLR}} \varpropto L^{\alpha}$ relationship \citep{2019FrASS...6...75P}. Because only $g$ and $r$ band data of ZTF are used, the redshift is limited to be smaller than 0.09. Most of the selected 23 AGNs are located in the middle of the R-L relation which has large dispersions. With more data released and the improved quality for ZTF i band lightcurves, as well as other large multi-epoch photometric sky surveys in the future, we expect to expand the ranges of redshift and luminosity of our sample.

Since the 23 AGNs have previous ${\rm H}\beta$ time lag measurements from SRM, we can compare our obtained ${\rm H}\alpha$ lags with them. Some previous works have found that the ${\rm H}\alpha$ time lags are slightly larger than the H$\beta$ time lags when measured at the same time \citep{2000ApJ...533..631K,2010ApJ...716..993B}. As shown in Figure 15, our result is consistent with this relation, which proves that our measured ${\rm H}\alpha$ time lags are mostly reasonable. By comparing different points in Figure 15 for List I, II and III, it can be found that the AGNs with consistent lag distributions of the three methods show small scatters. It can ensure our broadband PRM results furthermore. By comparing the results obtained with different methods in Figure 15, the ICCF-Cut results show the smallest scatters which means that the ICCF-Cut may be more suitable and convincing than the JAVELIN and $\chi^2$ methods for the broadband PRM. Moreover, the List I sample (except PG 0007+106) generally have less scatters on the ${\rm H}\alpha$-H$\beta$ lag distribution than those from List II and List III, since the photometric and spectroscopic data are synchronous. Some large scatters for List II and List III samples in Figure 15 may come from the asynchrony of the photometry and the spectra.

\section{Summary}

From AGNs at redshift $z<0.09$ in the ZTF DR16 catalog, we use the variability J index and the smoothness $s$ index to select AGNs with significant variabilities and high-quality lightcurves. From 3537 AGNs, we select 123 AGNs and obtain the H$\alpha$ time lags of 23 AGNs which all have the SRM lag results.

By assuming that the continuum flux in the $r$ band is equal to a fraction of the continuum flux in the $g$ band, we use the ICCF-Cut, JAVELIN and $\chi^2$ methods to calculate the ${\rm H}\alpha$ emission line time lags from the lightcurves in the $g$ and $r$ bands. For 4 out of 5 AGNs with simultaneous SRM observations in List I, we obtain the consistent results with the SRM results. For 8 AGNs in List II, the extracted H$\alpha$ lightcurves are similar to the lagged $g$ band lightcurves. For 6 out of the 8 AGNs, the lag distributions of the three methods are consistent. For the other 2 AGNs in List II, the lag distributions of ICCF-Cut and JAVELIN methods are consistent, while the $\chi^2$ lags are much larger. For 10 AGNs in List III, the inconsistency between the results obtained from three methods make the results not as convincing as in List I and II. We still need more observational data or improve the methods to determine the reliable H$\alpha$ lags.

We compare our derived ${\rm H}\alpha$ time lags with the ${\rm H}\beta$ $R-L$ relationship obtained from the SRM, and find that our results are consistent with previous works. By comparing the obtained H$\alpha$ lags with the SRM H$\beta$ lags, we can check the reliability of our H$\alpha$ lags furthermore. We noticed that the scatters of the ICCF-Cut results are smaller than those of the JAVELIN and $\chi^2$ methods, which means that the ICCF-Cut may be more suitable than the JAVELIN and $\chi^2$ methods for the broadband PRM. 
In general, the ${\rm H}\alpha$ lags obtained from the broadband PRM are slightly larger than or consistent with the ${\rm H}\beta$ lags obtained from SRM, which is consistent with previous works and theoretical predictions \citep[e.g.,][]{1988PASP..100.1041C, 1995ApJ...455L.119B, 2004ApJ...606..749K}.

All the results above indicate that after selecting AGNs with J index and $s$ index, combining the ICCF-Cut, JAVELIN and $\chi^2$ methods is an efficient way to obtain the reliable ${\rm H}\alpha$ line lags with the broadband PRM. These PRM methods can be used to study the BLR sizes and BH masses of a large sample of AGNs in the era of large multi-epoch photometric sky surveys \citep[e.g.,][]{2023A&A...675A.163C}, such as the Legacy Survey of Space and Time (LSST of the Vera C. Rubin Observatory) \citep{2017arXiv170804058L} and the multi-band photometric survey from the Wide Field Survey Telescope \citep[WFST;][]{2023SCPMA..6609512W}, in the near future. These surveys will improve the observation cadence (1-3 days), enabling the application of the PRM methods along with the modifications we have proposed.

$Acknowledgements$.
This work is supported by the National Key R\&D Program of China No. 2022YFF0503401. We are thankful for the support of the National Science Foundation of China (11927804 and 12133001). We acknowledge the science research grant from the China Manned Space Project with No. CMS-CSST-2021-A06. We acknowledge the support of the staff of the Xinglong 2.16m telescope. This work was partially Supported by the Open Project Program of the CAS Key Laboratory of Optical Astronomy, National Astronomical Observatories, Chinese Academy of Sciences. We acknowledge the archived data from ZTF, SDSS, and LAMOST.
This work is based on observations obtained with the Samuel Oschin Telescope 48-inch and the 60-inch Telescope at the Palomar Observatory as part of the Zwicky Transient Facility project. ZTF is supported by the National Science Foundation under Grant No. AST-2034437 and a collaboration including Caltech, IPAC, the Weizmann Institute for Science, the Oskar Klein Center at Stockholm University, the University of Maryland, Deutsches Elektronen-Synchrotron and Humboldt University, the TANGO Consortium of Taiwan, the University of Wisconsin at Milwaukee, Trinity College Dublin, Lawrence Livermore National Laboratories, and IN2P3, France. Operations are conducted by COO, IPAC, and UW.

\bibliography{ztf.bib}
\bibliographystyle{aasjournal}

\appendix

Here we present the spectra used in this paper (Figure A1), as well as the lightcurves and lag distributions of the AGNs in List II (Figure B1-B8) and  List III (Figure C1-C10). We also provide the data of the final lightcurves used for the delay estimations of 23 AGNs in electronic forms\footnote{\href{https://github.com/PhotoRM/ICCF-Cut/tree/main/lightcurves}{https://github.com/PhotoRM/ICCF$-$Cut/tree/main/lightcurves}}.

\renewcommand\thefigure{A\arabic{figure}}
\setcounter{figure}{0}
\begin{figure*}[h]
\centering
\includegraphics[height=0.95\textwidth]{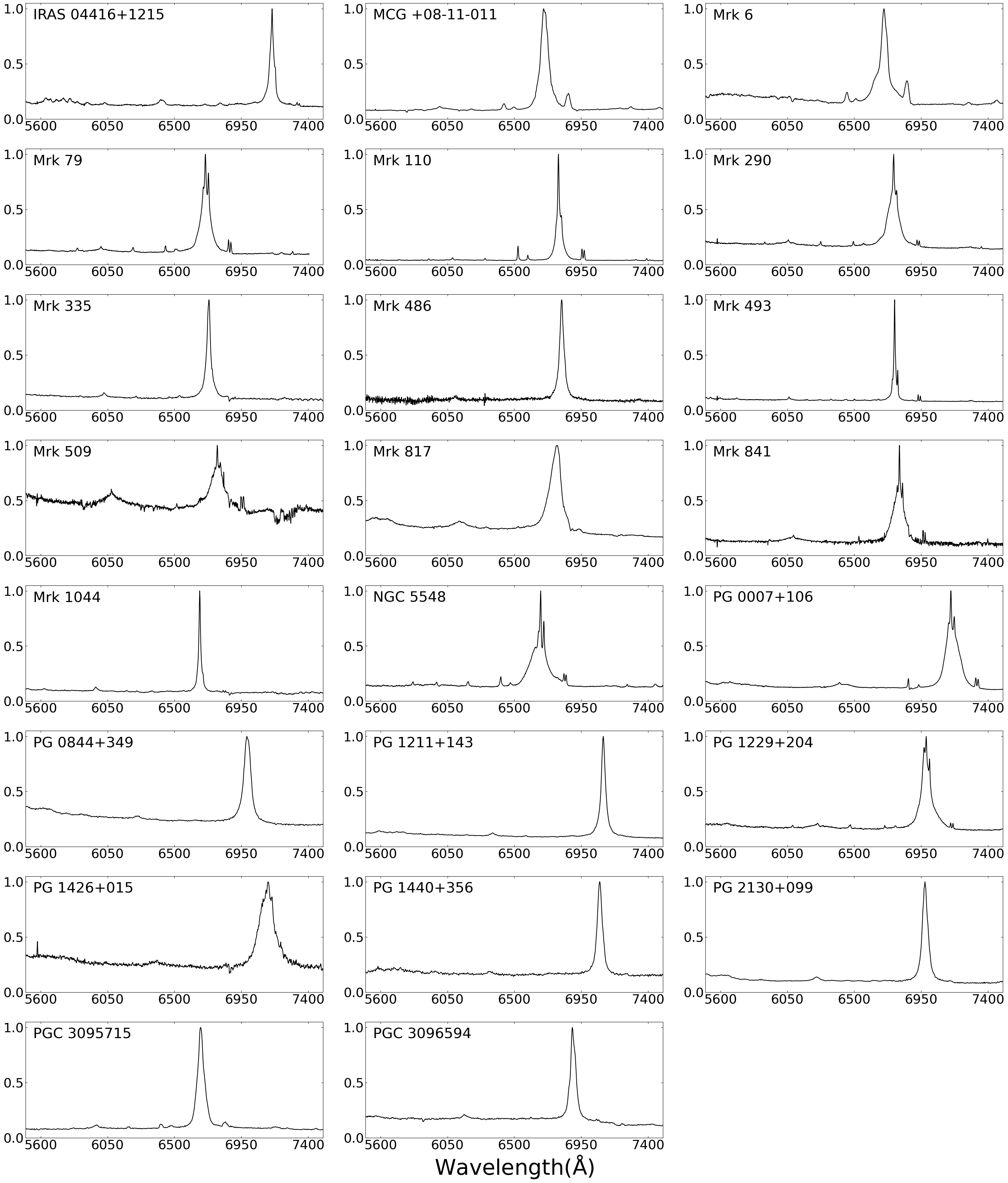}
\caption{Spectra of 23 AGNs used in this paper. The flux is normalized to the peak value of H$\alpha$ emission line. The wavelength range shown here refers to that of ZTF r band, which is used to calculate the H$\alpha$ ratio in r band (i.e., $F_{H\alpha}/F_{line}$ in Equation 7).}
\end{figure*}

\renewcommand\thefigure{B\arabic{figure}}
\setcounter{figure}{0}
\begin{figure*}[h]
\plotone{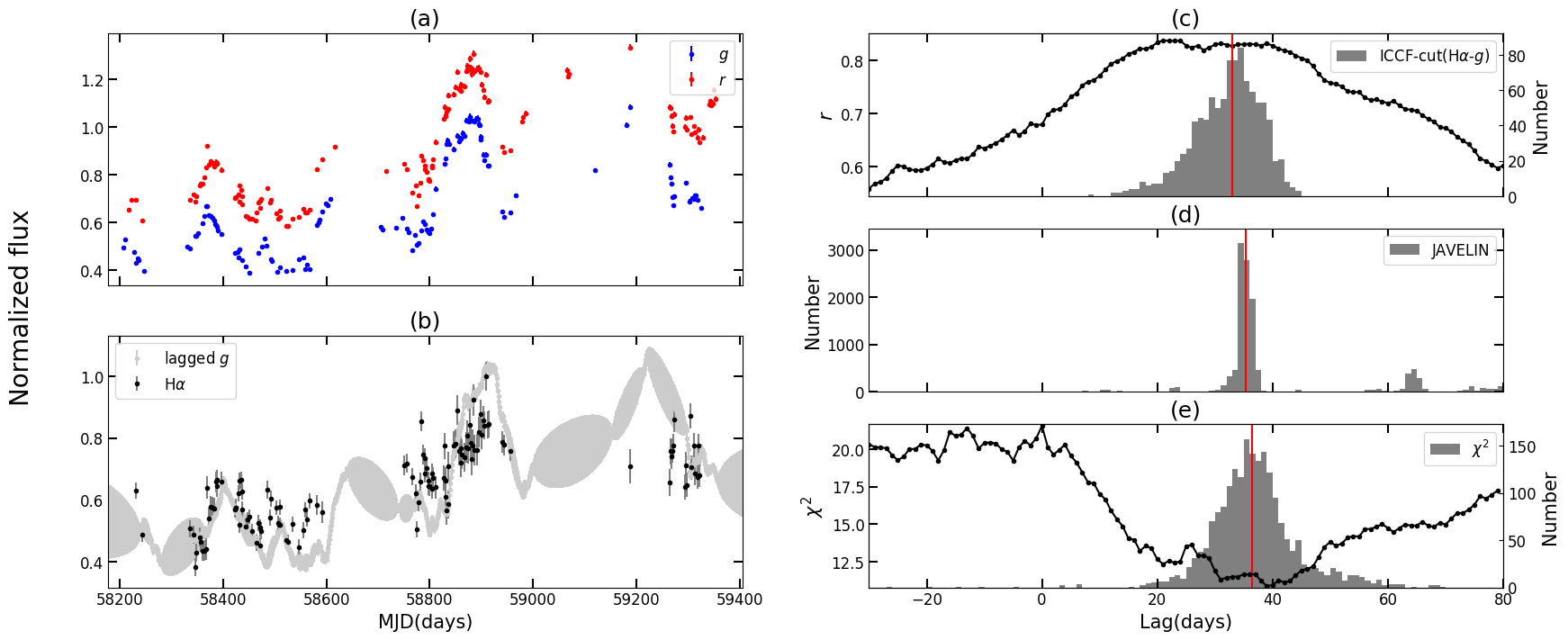}
\caption{Same as Figure 4 but for MCG +08-11-011.}
\end{figure*}

\begin{figure*}[h]
\plotone{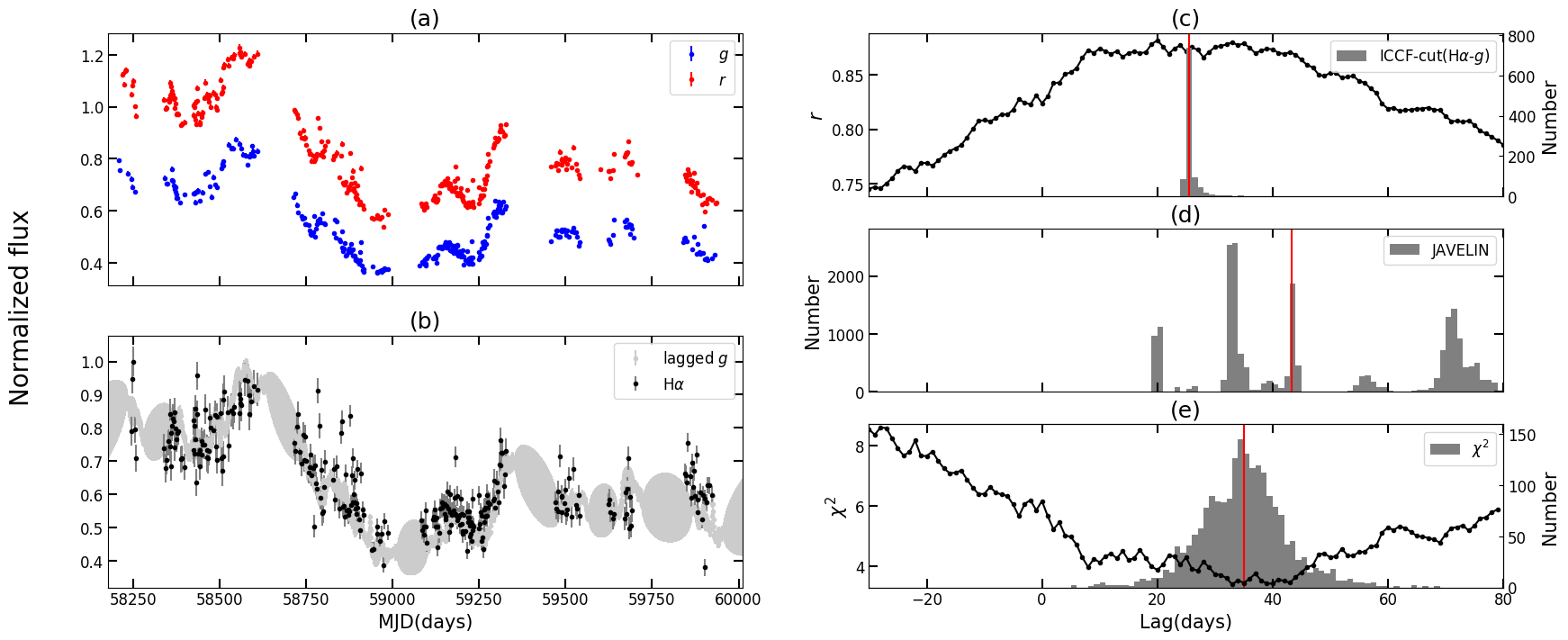}
\caption{Same as Figure 4 but for Mrk 6.}
\end{figure*}

\begin{figure*}[h]
\plotone{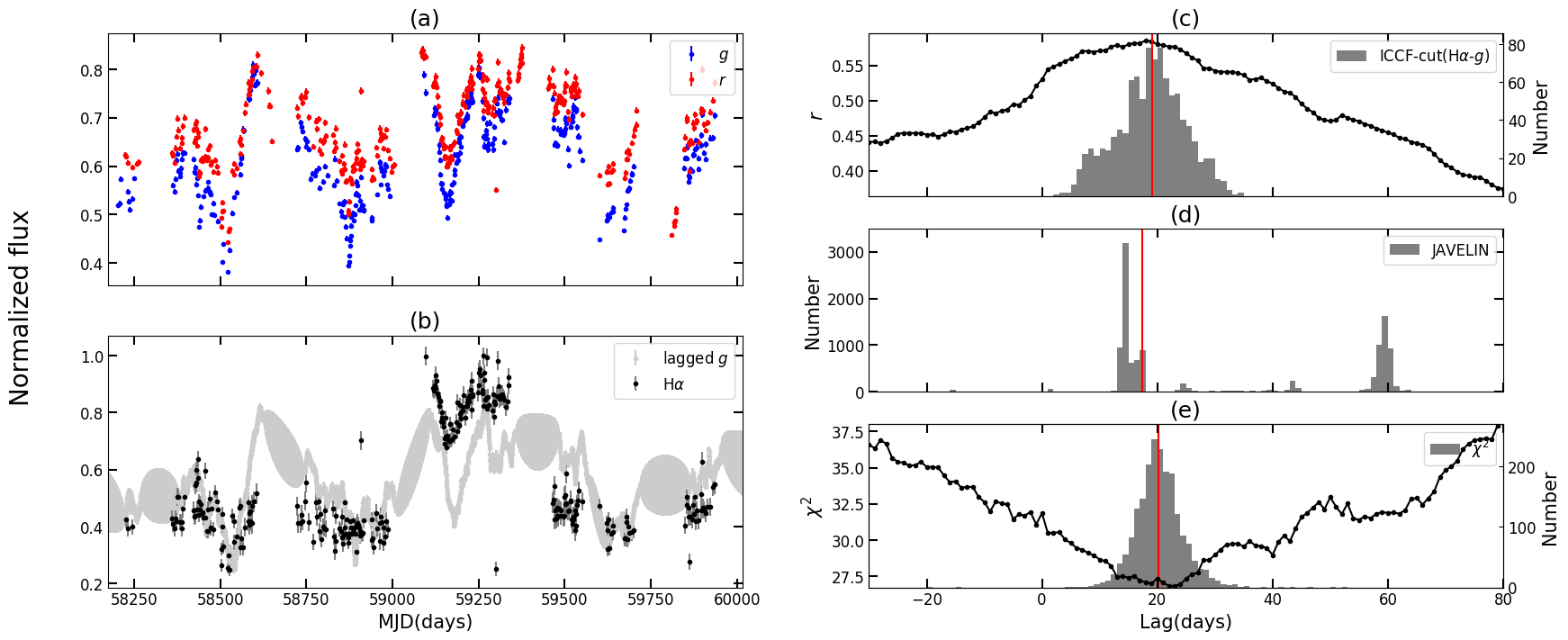}
\caption{Same as Figure 4 but for Mrk 79.}
\end{figure*}

\begin{figure*}[h]
\plotone{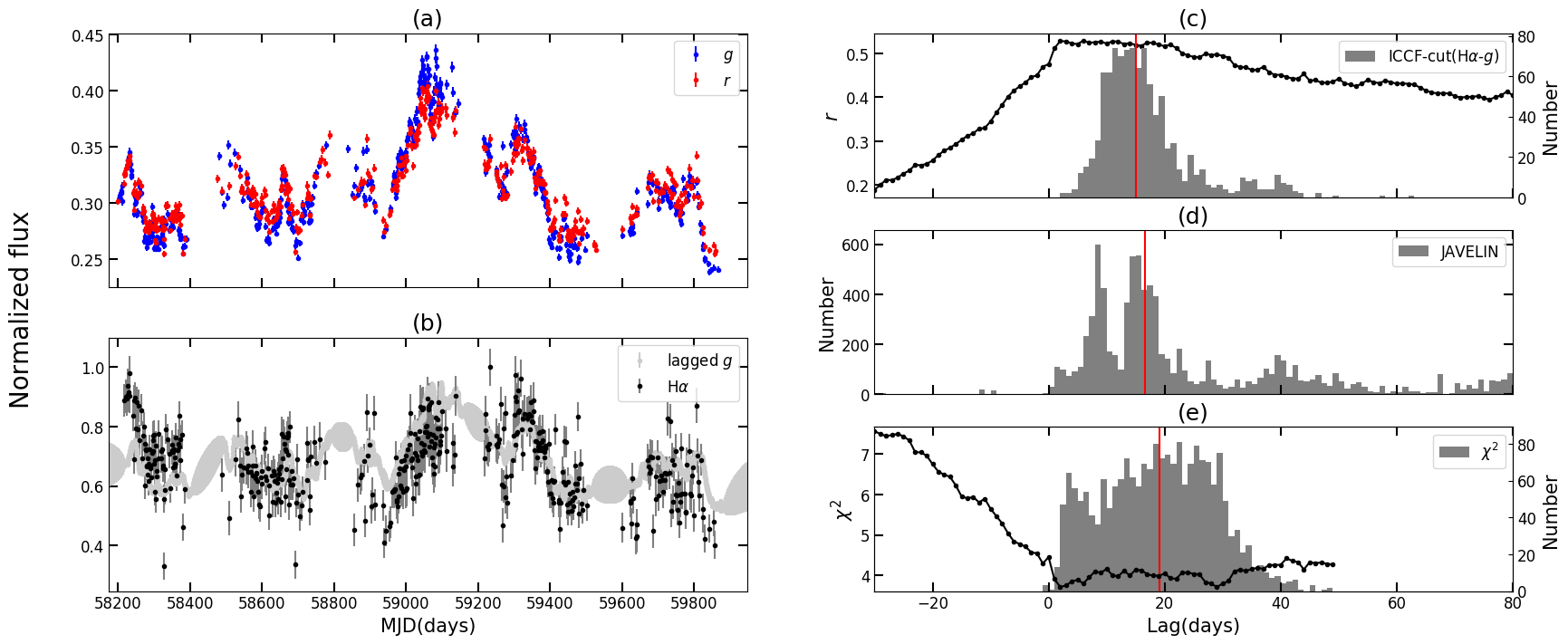}
\caption{Same as Figure 4 but for Mrk 493.}
\end{figure*}

\begin{figure*}[h]
\plotone{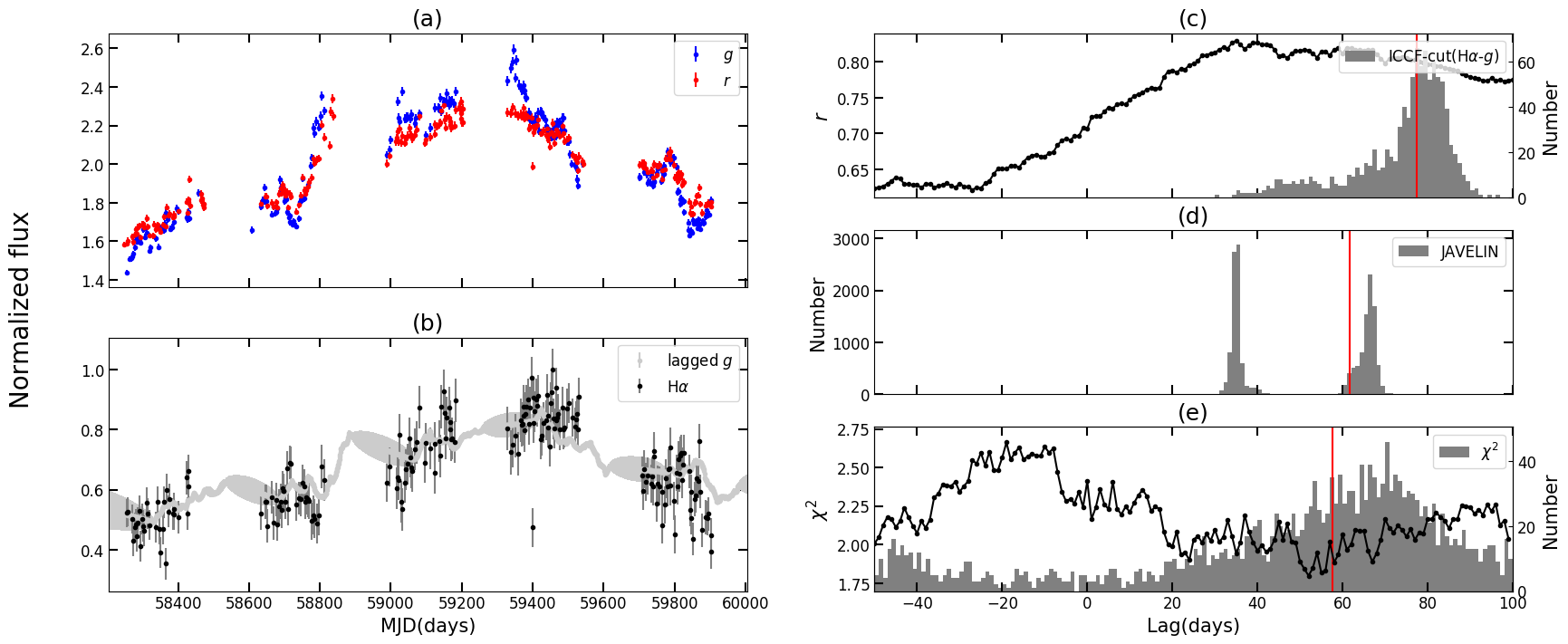}
\caption{Same as Figure 4 but for Mrk 509.}
\end{figure*}

\begin{figure*}[h]
\plotone{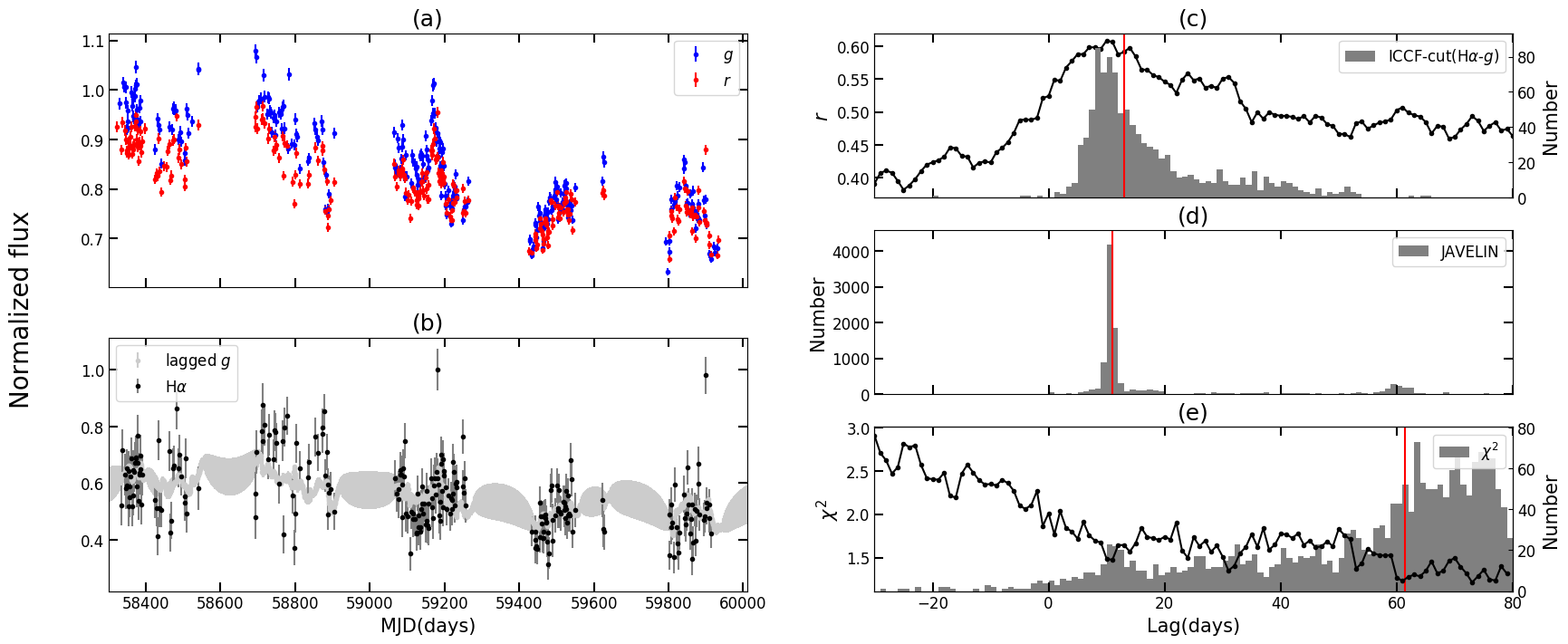}
\caption{Same as Figure 4 but for Mrk 1044.}
\end{figure*}

\begin{figure*}[h]
\plotone{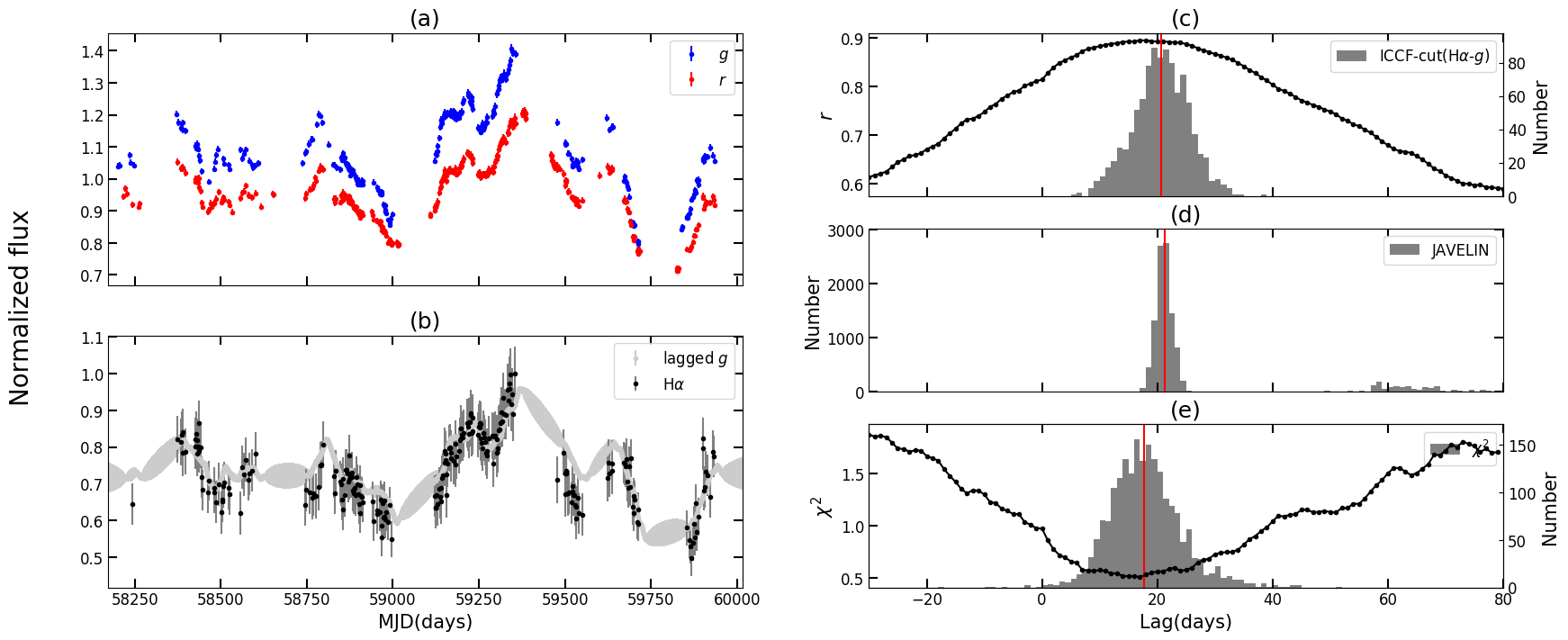}
\caption{Same as Figure 4 but for PG 0844+349.}
\end{figure*}

\begin{figure*}[h]
\plotone{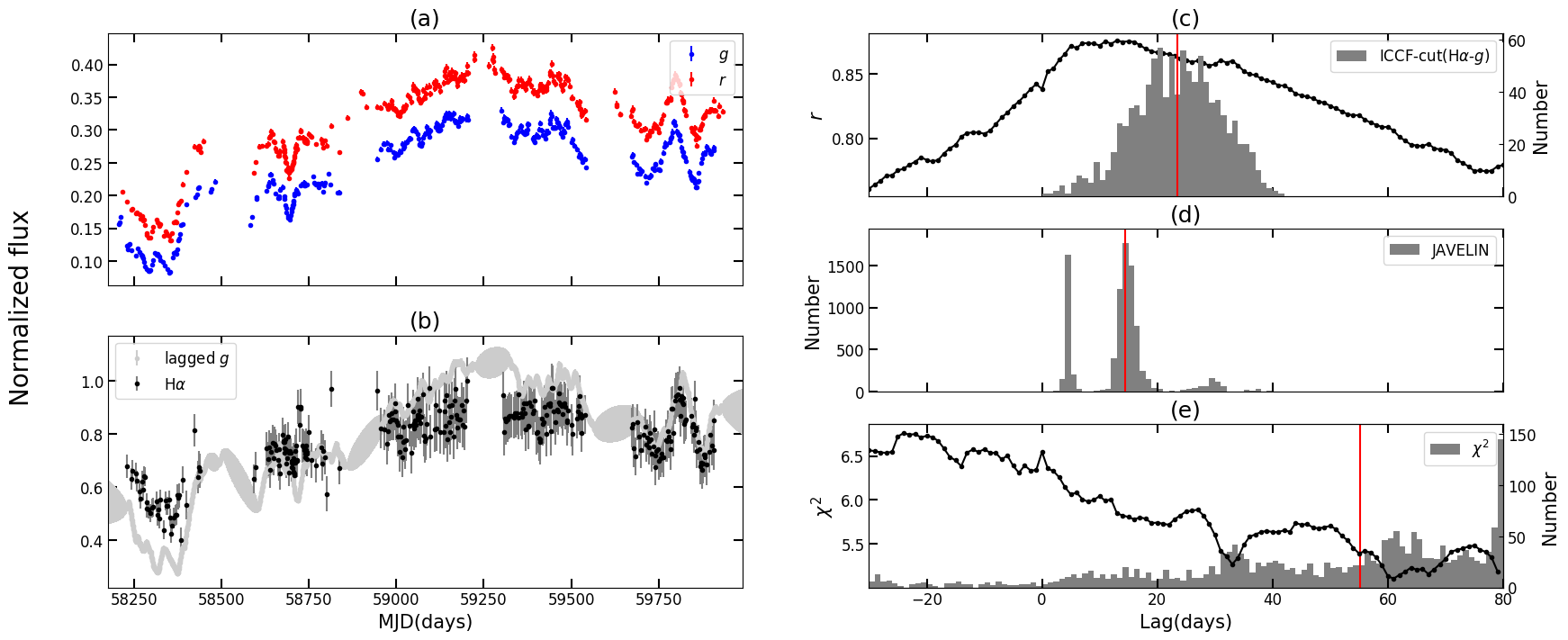}
\caption{Same as Figure 4 but for PGC 3096594.}
\end{figure*}

\renewcommand\thefigure{C\arabic{figure}}
\setcounter{figure}{0}
\begin{figure*}[h]
\plotone{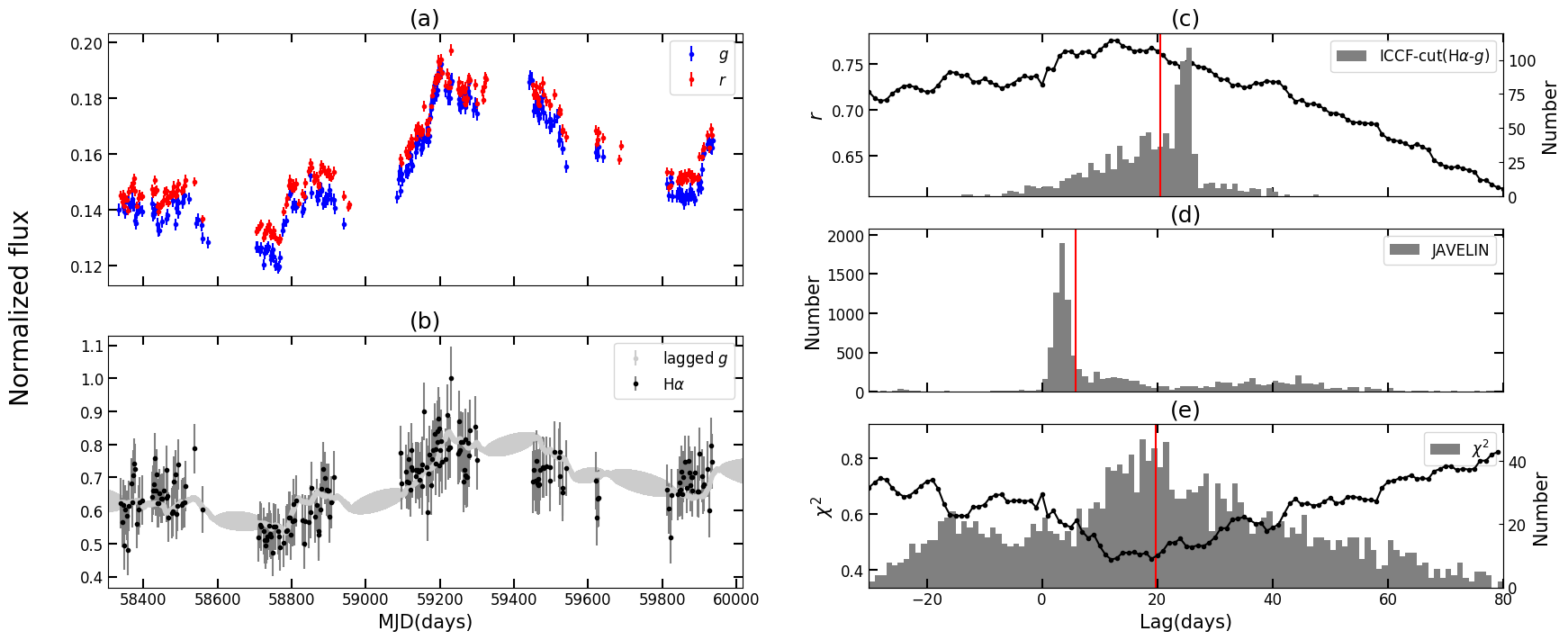}
\caption{Same as Figure 4 but for IRAS 04416+1215.}
\end{figure*}

\begin{figure*}[h]
\plotone{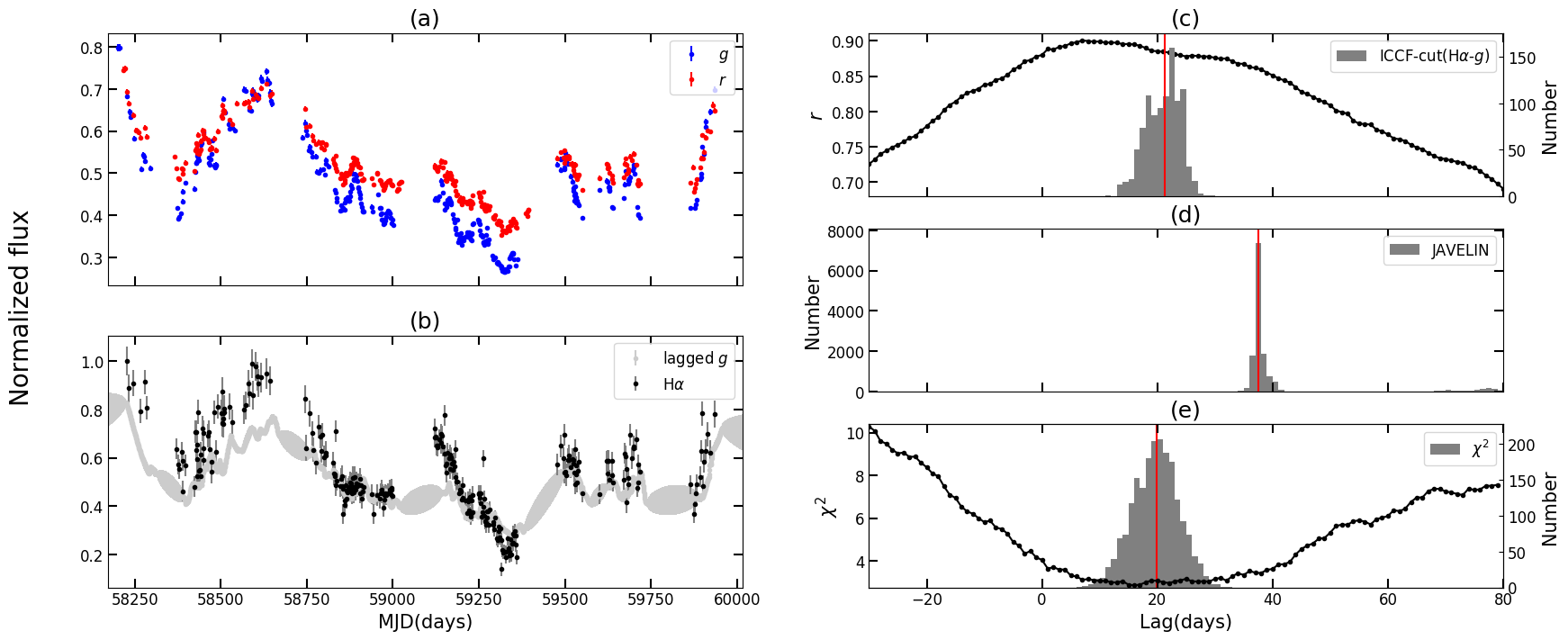}
\caption{Same as Figure 4 but for Mrk 110.}
\end{figure*}

\begin{figure*}[h]
\plotone{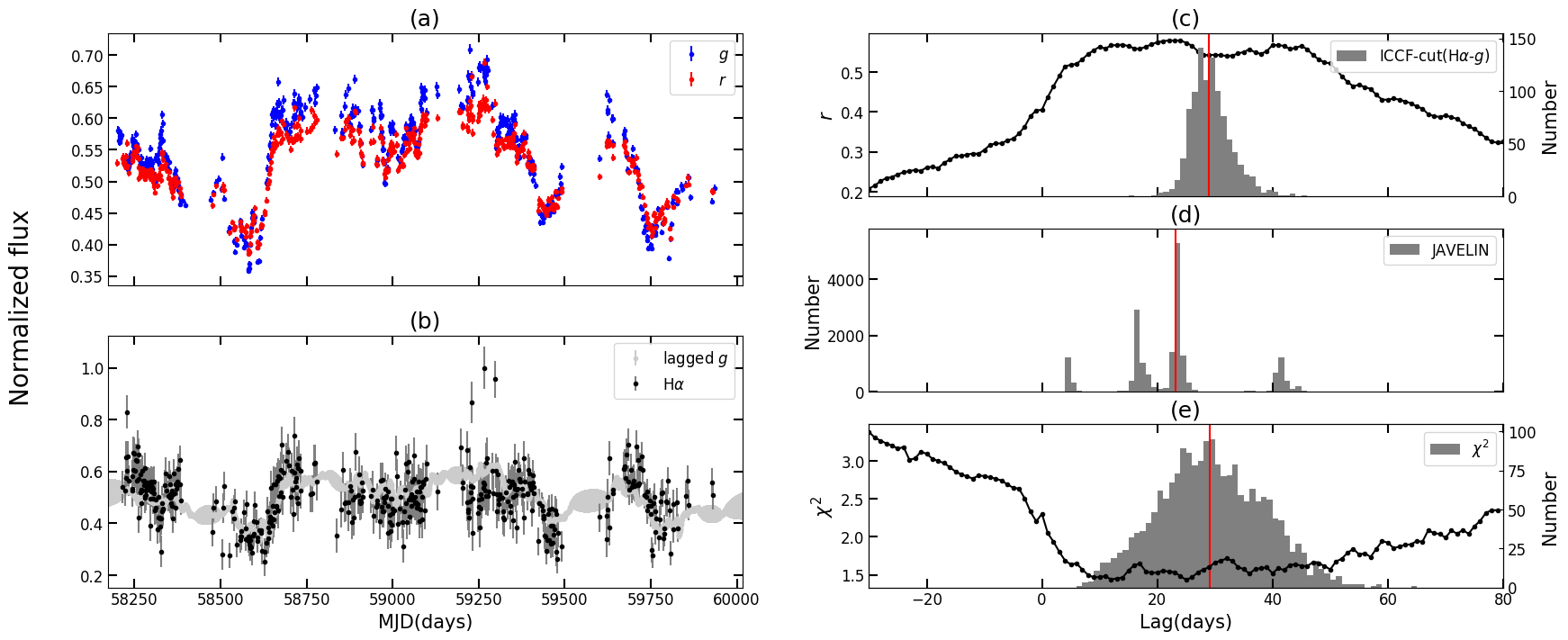}
\caption{Same as Figure 4 but for Mrk 290.}
\end{figure*}

\begin{figure*}[h]
\plotone{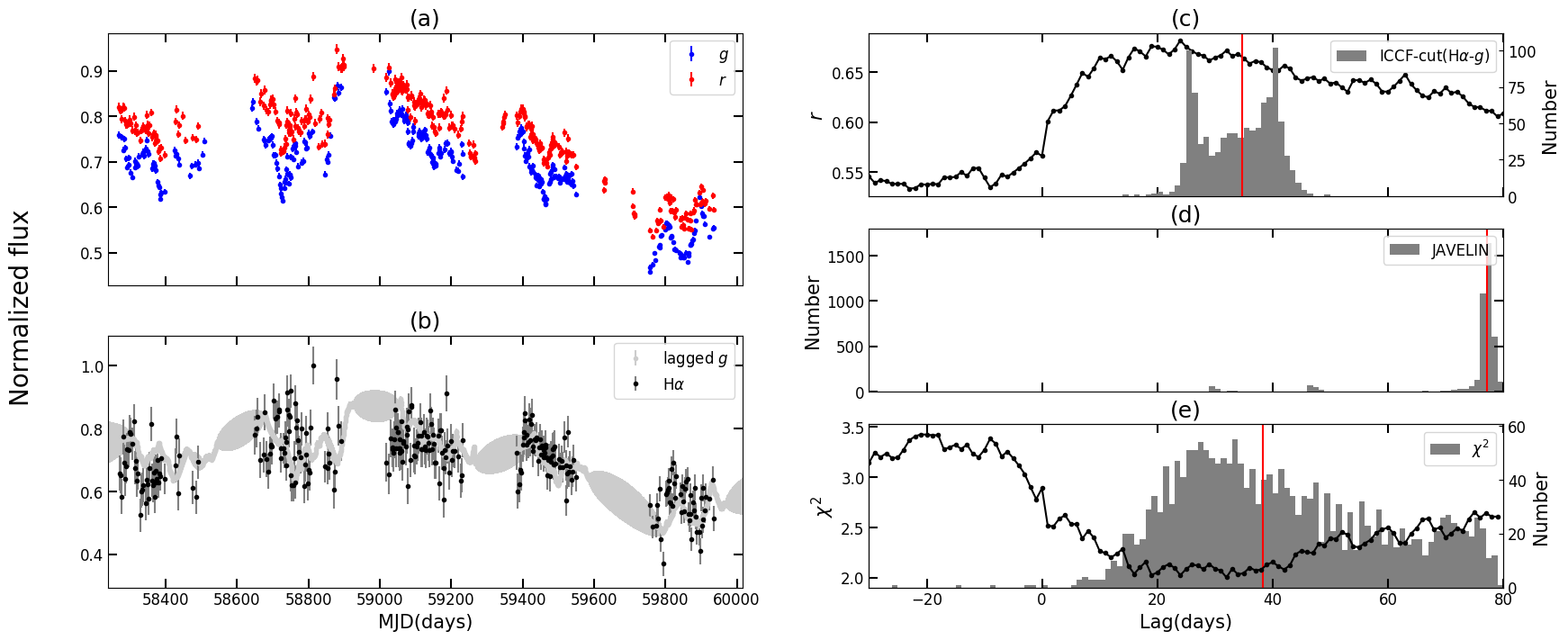}
\caption{Same as Figure 4 but for Mrk 335.}
\end{figure*}

\begin{figure*}[h]
\plotone{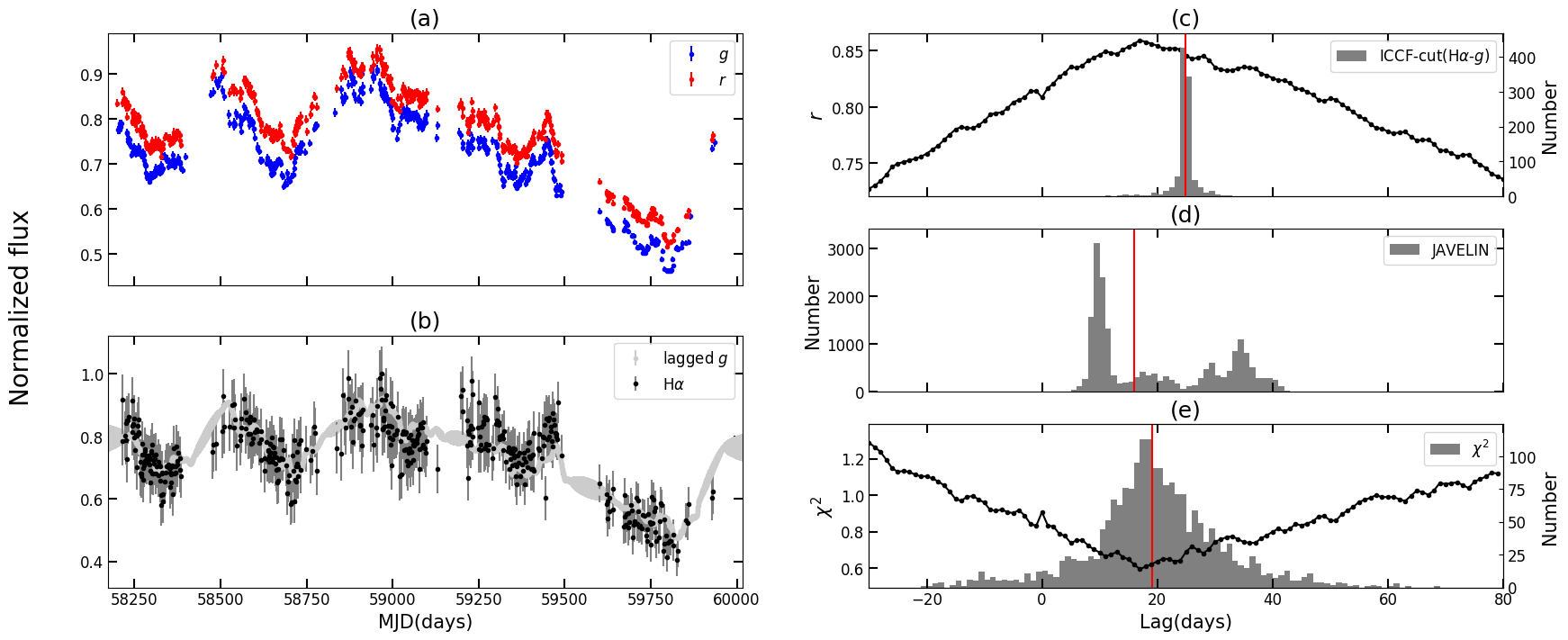}
\caption{Same as Figure 4 but for Mrk 486.}
\end{figure*}

\begin{figure*}[h]
\plotone{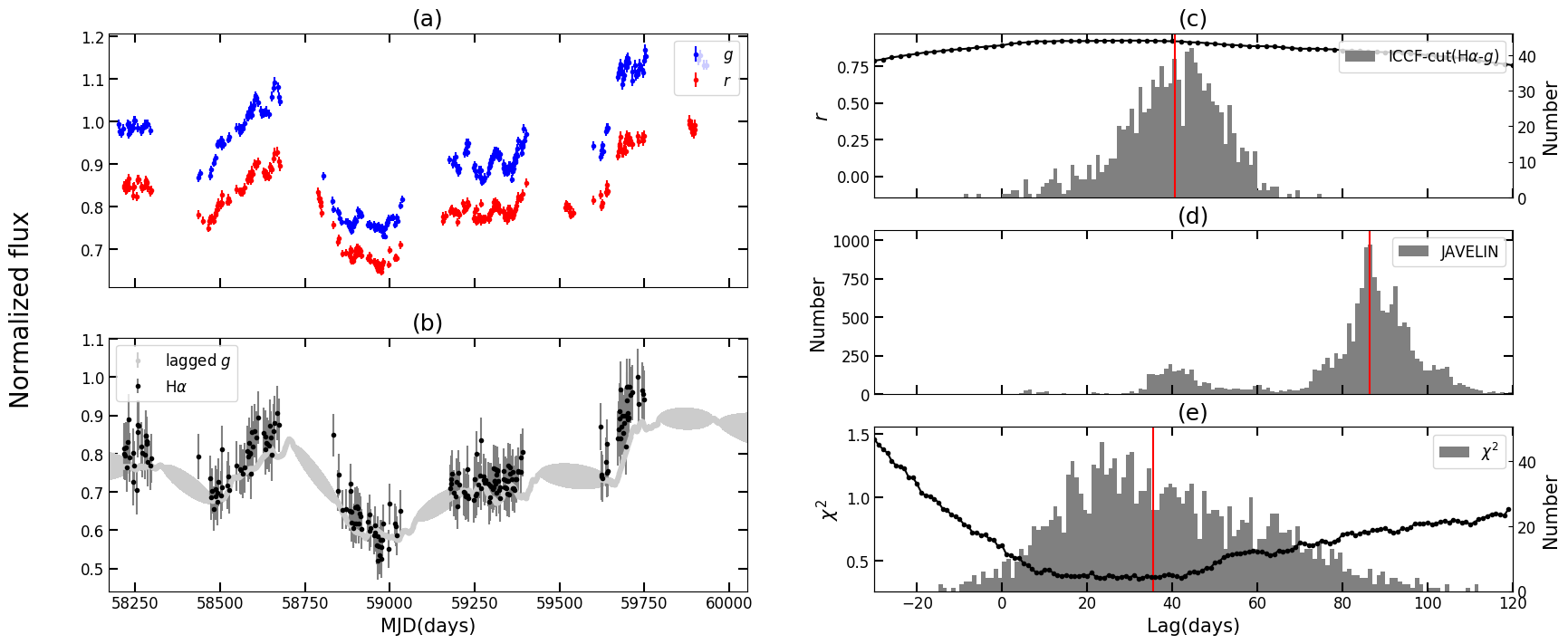}
\caption{Same as Figure 4 but for PG 1211+143.}
\end{figure*}

\begin{figure*}[h]
\plotone{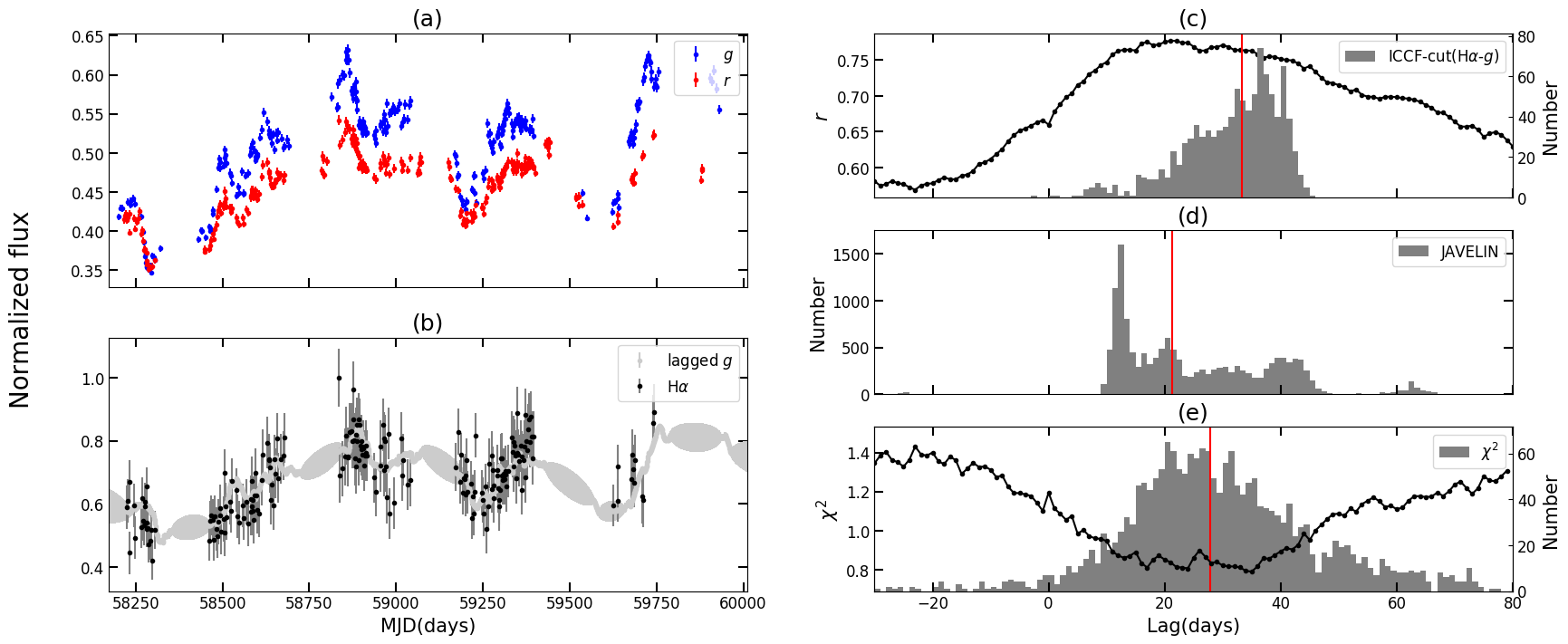}
\caption{Same as Figure 4 but for PG 1229+204.}
\end{figure*}

\begin{figure*}[h]
\plotone{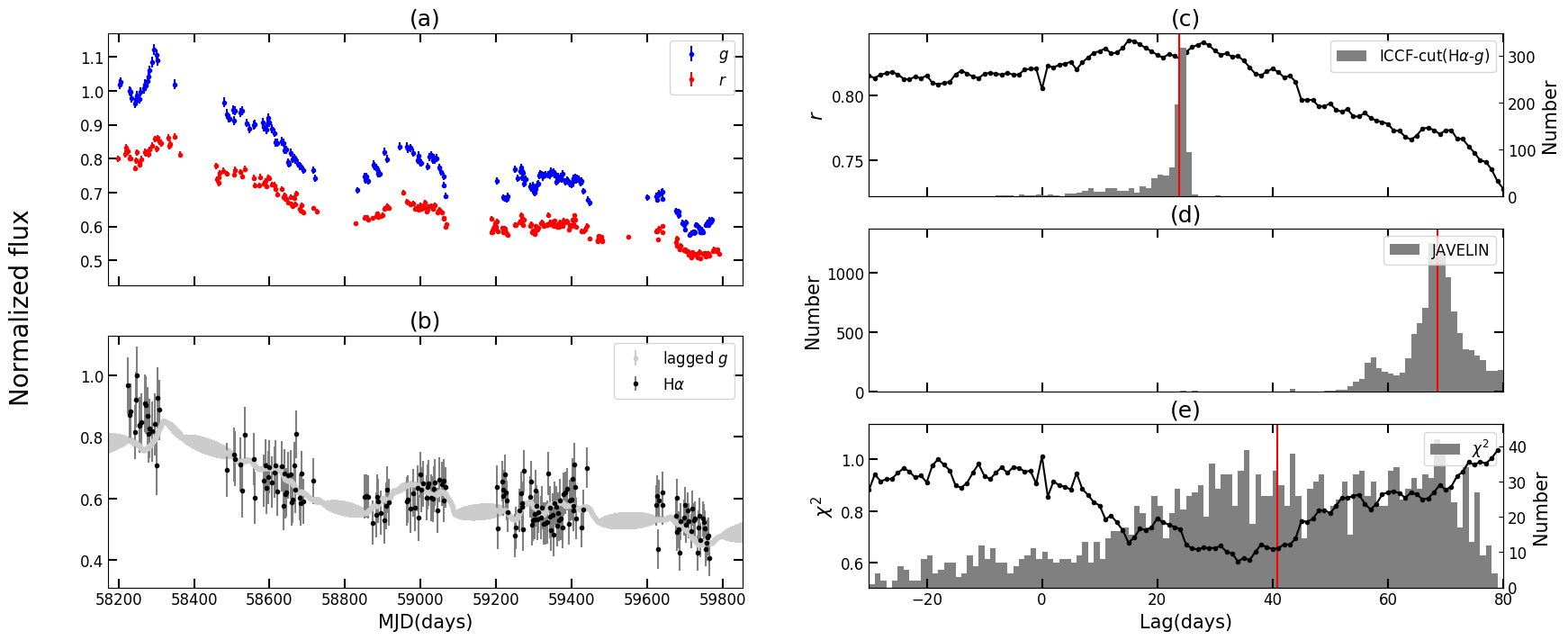}
\caption{Same as Figure 4 but for PG 1426+015.}
\end{figure*}

\begin{figure*}[h]
\plotone{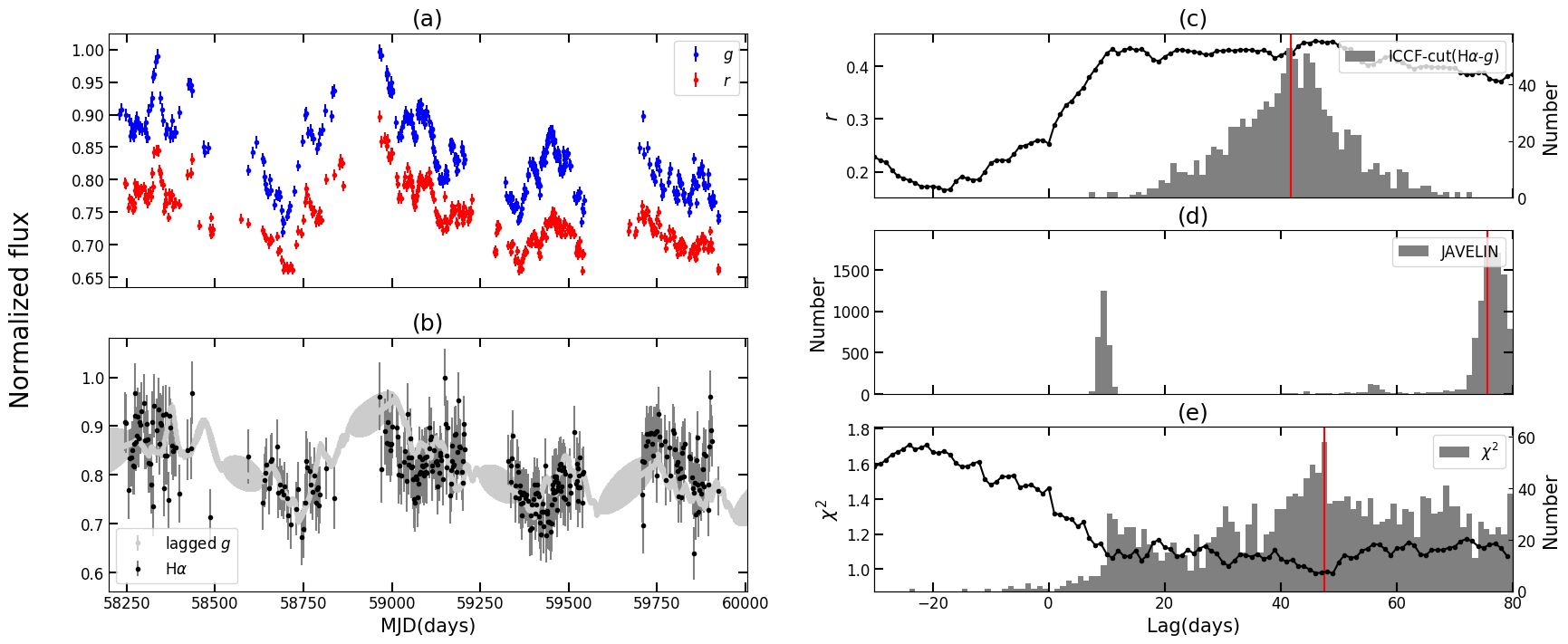}
\caption{Same as Figure 4 but for PG 2130+099.}
\end{figure*}

\begin{figure*}[h]
\plotone{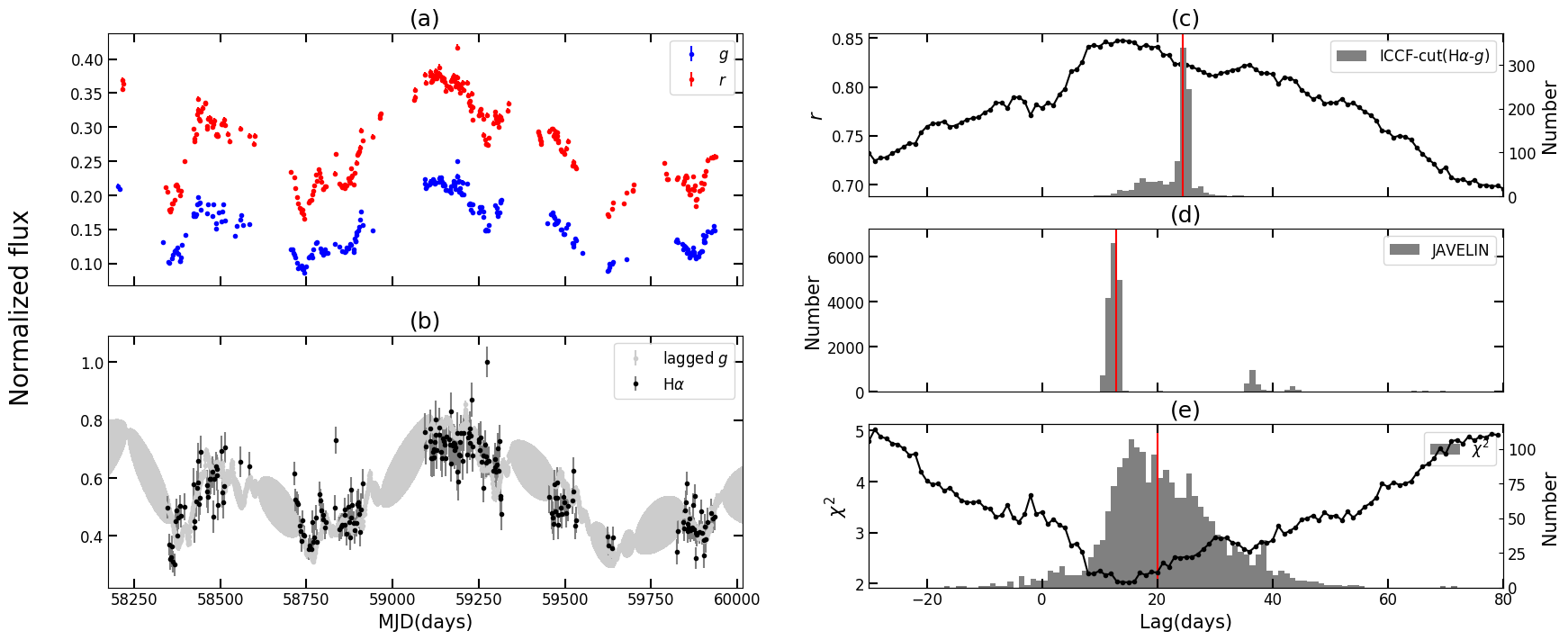}
\caption{Same as Figure 4 but for PGC 3095715.}
\end{figure*}

\end{document}